
\def\cft           {conformal field theory}
\def\cfts          {conformal field theories}
\def\eE            {{\rm e}}
\def\epss          {\epsilon_\sigma}
\def\vareps        {\epsilon}
\def\findim        {finite-dimen\-sional}
\def\gagr          {Galois group}
\def\Gal           {{\cal G}\!a\ell}
\def\gallq         {\Gal(L/\rationals)}
\def\galml         {\Gal(M/L)}
\def\galmq         {\Gal(M/\rationals)}
\def\gb            {\bar{\rm g}}
\def\ii            {{\rm i}}
\def\Mskip         {\vskip1mm}
\def\one           {{\bf 1}}
\def\onedim        {one-dimensional}
\def\P             {{\cal P}}
\def\Q             {{\cal Q}}
\def\qdim          {quantum dimension}
\def\qzn           {\rationals(\zeta_n)}
\def\rationals     {{\bf Q}}
\def\sigmA         {\dot\sigma}
\def\sigmal        {\sigma_{(\ell)}^{}}
\def\sigmAl        {\dot\sigma_{(\ell)}^{}}

\def\sigmaL        {\sigma_L^{}}
\def\sigmaM        {\sigma_M^{}}

\def\smat          {$S$-matrix}
\def\SO            {{\rm so}}
\def\SU            {{\rm su}}
\def\sumik         {\sum_{k\in I}}
\def\sumil         {\sum_{\ell\in I}}
\def\wh            {\hat w}
\def\wzwt          {WZW theory}

\def\zett          {{\bf Z}}
 \def\ChapGal{{2}} \def\ChapInf{{3}} \def\ChapNew{{4}}
\def\ChapPure{{5}} \def\ChapExt{{6}} 

\input phyzzx
\catcode`@=11 
\def\space@ver#1{\let\@sf=\empty \ifmmode #1\else \ifhmode
   \edef\@sf{\spacefactor=\the\spacefactor}\unskip${}#1$\relax\fi\fi}
\def\attach#1{\space@ver{\strut^{\mkern 2mu #1} }\@sf\ }
\newtoks\foottokens
%
%
%
%
\newbox\leftpage \newdimen\fullhsize \newdimen\hstitle \newdimen\hsbody
\newif\ifreduce  \reducefalse
\def\almostshipout#1{\if L\lr \count2=1
      \global\setbox\leftpage=#1 \global\let\lr=R
  \else \count2=2
    \shipout\vbox{\special{dvitops: landscape}
      \hbox to\fullhsize{\box\leftpage\hfil#1}} \global\let\lr=L\fi}
\def\smallsize{\relax
\font\eightrm=cmr8 \font\eightbf=cmbx8 \font\eighti=cmmi8
\font\eightsy=cmsy8 \font\eightsl=cmsl8 \font\eightit=cmti8
\font\eightt=cmtt8
\def\eightpoint{\relax
\textfont0=\eightrm  \scriptfont0=\sixrm
\scriptscriptfont0=\sixrm
\def\rm{\fam0 \eightrm \f@ntkey=0}\relax
\textfont1=\eighti  \scriptfont1=\sixi
\scriptscriptfont1=\sixi
\def\oldstyle{\fam1 \eighti \f@ntkey=1}\relax
\textfont2=\eightsy  \scriptfont2=\sixsy
\scriptscriptfont2=\sixsy
\textfont3=\tenex  \scriptfont3=\tenex
\scriptscriptfont3=\tenex
\def\it{\fam\itfam \eightit \f@ntkey=4 }\textfont\itfam=\eightit
\def\sl{\fam\slfam \eightsl \f@ntkey=5 }\textfont\slfam=\eightsl
\def\bf{\fam\bffam \eightbf \f@ntkey=6 }\textfont\bffam=\eightbf
\scriptfont\bffam=\sixbf   \scriptscriptfont\bffam=\sixbf
\def\tt{\fam\ttfam \eightt \f@ntkey=7 }
\def\caps{\fam\cpfam \tencp \f@ntkey=8 }\textfont\cpfam=\tencp
\setbox\strutbox=\hbox{\vrule height 7.35pt depth 3.02pt width\z@}
\samef@nt}
\def\Eightpoint{\eightpoint \relax
  \ifsingl@\subspaces@t2:5;\else\subspaces@t3:5;\fi
  \ifdoubl@ \multiply\baselineskip by 5
            \divide\baselineskip by 4\fi }
\parindent=16.67pt
\itemsize=25pt
\thinmuskip=2.5mu
\medmuskip=3.33mu plus 1.67mu minus 3.33mu
\thickmuskip=4.17mu plus 4.17mu
\def\thinspace{\kern .13889em }
\def\negthinspace{\kern-.13889em }
\def\enspace{\kern.416667em }
\def\enskip{\hskip.416667em\relax}
\def\quad{\hskip.83333em\relax}
\def\qquad{\hskip1.66667em\relax}
\def\crr{\cropen{8.3333pt}}
\foottokens={\Eightpoint\singlespace}
\def\papersize{\SIZE\OFFSET\skip\footins=\bigskipamount}
\def\SIZE{\hsize=11.8truecm\vsize=17.5truecm}
\def\OFFSET{\voffset=-1.3truecm\hoffset=  .14truecm}
\message{STANDARD CERN-PREPRINT FORMAT}
\def\attach##1{\space@ver{\strut^{\mkern 1.6667mu ##1} }\@sf\ }
\def\PH@SR@V{\doubl@true\baselineskip=20.08pt plus .1667pt minus .0833pt
             \parskip = 2.5pt plus 1.6667pt minus .8333pt }
\def\author##1{\vskip\frontpageskip\titlestyle{\tencp ##1}\nobreak}
\def\address##1{\par\kern 4.16667pt\titlestyle{\tenpoint\it ##1}}
\def\andaddress{\par\kern 4.16667pt \centerline{\sl and} \address}
\def\abstract{\vskip2\frontpageskip\centerline{\tenrm Abstract}
              \vskip\headskip }
\def\cases##1{\left\{\,\vcenter{\Tenpoint\m@th
    \ialign{$####\hfil$&\quad####\hfil\crcr##1\crcr}}\right.}
\def\matrix##1{\,\vcenter{\Tenpoint\m@th
    \ialign{\hfil$####$\hfil&&\quad\hfil$####$\hfil\crcr
      \mathstrut\crcr\noalign{\kern-\baselineskip}
     ##1\crcr\mathstrut\crcr\noalign{\kern-\baselineskip}}}\,}
\Tenpoint
}
\def\Smallsize{\smallsize\reducetrue
\let\lr=L
\hstitle=8truein\hsbody=4.75truein\fullhsize=24.6truecm\hsize=\hsbody
\output={
  \almostshipout{\leftline{\vbox{\makeheadline
  \pagebody\makefootline}}}\advancepageno
     }
\special{dvitops: landscape}
\def\makeheadline{
\iffrontpage\line{\the\headline}
             \else\vskip .0truecm\line{\the\headline}\vskip .5truecm \fi}
\def\makefootline{\iffrontpage\vskip  0.truecm\line{\the\footline}
               \vskip -.15truecm\line{\the\date\hfil}
              \else\line{\the\footline}\fi}
\paperheadline={
\iffrontpage\hfil
               \else
               \tenrm\hss $-$\ \folio\ $-$\hss\fi    }
\paperstyle}
%
%
%
%
%
%
%
%
%
\newcount\referencecount     \referencecount=0
\newif\ifreferenceopen       \newwrite\referencewrite
\newtoks\rw@toks
\def\NPrefmark#1{\attach{\scriptscriptstyle [ #1 ] }}
\let\PRrefmark=\attach
\def\refmark#1{\relax\ifPhysRev\PRrefmark{#1}\else\NPrefmark{#1}\fi}
\def\refend{\refmark{\number\referencecount}}
\newcount\lastrefsbegincount \lastrefsbegincount=0
\def\refsend{\refmark{\count255=\referencecount
   \advance\count255 by-\lastrefsbegincount
   \ifcase\count255 \number\referencecount
   \or \number\lastrefsbegincount,\number\referencecount
   \else \number\lastrefsbegincount-\number\referencecount \fi}}
\def\refch@ck{\chardef\rw@write=\referencewrite
   \ifreferenceopen \else \referenceopentrue
   \immediate\openout\referencewrite=referenc.texauxil \fi}
%
{\catcode`\^^M=\active 
  \gdef\obeyendofline{\catcode`\^^M\active \let^^M\ }}%
%
{\catcode`\^^M=\active 
  \gdef\ignoreendofline{\catcode`\^^M=5}}
{\obeyendofline\gdef\rw@start#1{\def\t@st{#1} \ifx\t@st\blankend%
\endgroup \@sf \relax \else \ifx\t@st\bl@nkend \endgroup \@sf \relax%
\else \rw@begin#1
\backtotext
\fi \fi } }
{\obeyendofline\gdef\rw@begin#1
{\def\n@xt{#1}\rw@toks={#1}\relax%
\rw@next}}
\def\blankend{}
{\obeylines\gdef\bl@nkend{
}}
\newif\iffirstrefline  \firstreflinetrue
\def\rwr@teswitch{\ifx\n@xt\blankend \let\n@xt=\rw@begin %
 \else\iffirstrefline \global\firstreflinefalse%
\immediate\write\rw@write{\noexpand\obeyendofline \the\rw@toks}%
\let\n@xt=\rw@begin%
      \else\ifx\n@xt\rw@@d \def\n@xt{\immediate\write\rw@write{%
        \noexpand\ignoreendofline}\endgroup \@sf}%
             \else \immediate\write\rw@write{\the\rw@toks}%
             \let\n@xt=\rw@begin\fi\fi \fi}
\def\rw@next{\rwr@teswitch\n@xt}
\def\rw@@d{\backtotext} \let\rw@end=\relax
\let\backtotext=\relax

\newdimen\refindent     \refindent=30pt
\def\refitem#1{\par \hangafter=0 \hangindent=\refindent \Textindent{#1}}
\def\REFNUM#1{\space@ver{}\refch@ck \firstreflinetrue%
 \global\advance\referencecount by 1 \xdef#1{\the\referencecount}}
\def\refnum#1{\space@ver{}\refch@ck \firstreflinetrue%
 \global\advance\referencecount by 1 \xdef#1{\the\referencecount}\refend}

\def\REF#1{\REFNUM#1%
 \immediate\write\referencewrite{%
 \noexpand\refitem{#1.}}%
\begingroup\obeyendofline\rw@start}
\def\ref{\refnum\?%
 \immediate\write\referencewrite{\noexpand\refitem{\?.}}%
\begingroup\obeyendofline\rw@start}
\def\Ref#1{\refnum#1%
 \immediate\write\referencewrite{\noexpand\refitem{#1.}}%
\begingroup\obeyendofline\rw@start}
\def\REFS#1{\REFNUM#1\global\lastrefsbegincount=\referencecount
\immediate\write\referencewrite{\noexpand\refitem{#1.}}%
\begingroup\obeyendofline\rw@start}
\newif\ifmref  
\newif\iffref  
\def\xrefsend{\xrefmark{\count255=\referencecount
\advance\count255 by-\lastrefsbegincount
\ifcase\count255 \number\referencecount
\or \number\lastrefsbegincount,\number\referencecount
\else \number\lastrefsbegincount-\number\referencecount \fi}}
\def\xrefsdub{\xrefmark{\count255=\referencecount
\advance\count255 by-\lastrefsbegincount
\ifcase\count255 \number\referencecount
\or \number\lastrefsbegincount,\number\referencecount
\else \number\lastrefsbegincount,\number\referencecount \fi}}
\def\xREFNUM#1{\space@ver{}\refch@ck\firstreflinetrue%
\global\advance\referencecount by 1
\xdef#1{\xrefend}}
\def\xrefend{\xrefmark{\number\referencecount}}
\def\xrefmark#1{[{#1}]}
\def\xRef#1{\xREFNUM#1\immediate\write\referencewrite%
{\noexpand\refitem{#1 }}\begingroup\obeyendofline\rw@start}%
\def\xREFS#1{\xREFNUM#1\global\lastrefsbegincount=\referencecount%
\immediate\write\referencewrite{\noexpand\refitem{#1 }}%
\begingroup\obeyendofline\rw@start}
\def\rrr#1#2{\relax\ifmref{\iffref\xREFS#1{#2}%
\else\xRef#1{#2}\fi}\else\xRef#1{#2}\xrefend\fi}
\def\multref#1#2{\mreftrue\freftrue{#1}%
\freffalse{#2}\mreffalse\xrefsend}
\referencecount=0
\def\refbreak{\hfil\penalty200\hfilneg}
\def\paperstyle{\papers}
\paperstyle   
%
%
%
\def\slacpub{\afterassignment\slacp@b\toks@}
\def\slacp@b{\edef\n@xt{\Pubnum={NIKHEF--H/\the\toks@}}\n@xt}
\let\pubnum=\slacpub
\expandafter\ifx\csname eightrm\endcsname\relax
    \let\eightrm=\ninerm \let\eightbf=\ninebf \fi

\font\seventeencp=cmcsc10 scaled\magstep3

\newif\ifCONF \CONFfalse
\newif\ifBREAK \BREAKfalse
\newif\ifsectionskip \sectionskiptrue

%
%
%
%
\def\NuclPhysProc{
\let\lr=L
\hstitle=8truein\hsbody=4.75truein\fullhsize=21.5truecm\hsize=\hsbody
\hstitle=8truein\hsbody=4.75truein\fullhsize=20.7truecm\hsize=\hsbody
\output={
  \almostshipout{\leftline{\vbox{\makeheadline
  \pagebody\makefootline}}}\advancepageno
     }
\def\papersize{\SIZE\OFFSET\skip\footins=\bigskipamount}
\def\SIZE{\hsize=10.0truecm\vsize=27.0truecm}
\def\OFFSET{\voffset=-1.4truecm\hoffset=-2.40truecm}
\message{NUCLEAR PHYSICS PROCEEDINGS FORMAT}
\def\makeheadline{
\iffrontpage\line{\the\headline}
             \else\vskip .0truecm\line{\the\headline}\vskip .5truecm \fi}
\def\makefootline{\iffrontpage\vskip  0.truecm\line{\the\footline}
               \vskip -.15truecm\line{\the\date\hfil}
              \else\line{\the\footline}\fi}
\paperheadline={\hfil}
\paperstyle}
%
%
%
%

%
%
%
%

%
%
%
%
\def\ReprintVolume{\smallsize
\def\papersize{\hsize=18.0truecm\vsize=23.1truecm\voffset -.73truecm
    \hoffset -.65truecm\skip\footins=\bigskipamount
    \normaldisplayskip= 20pt plus 5pt minus 10pt}
\message{REPRINT VOLUME FORMAT}
\paperstyle\baselineskip=.425truecm\parskip=0truecm
\def\makeheadline{
\iffrontpage\line{\the\headline}
             \else\vskip .0truecm\line{\the\headline}\vskip .5truecm \fi}
\def\makefootline{\iffrontpage\vskip  0.truecm\line{\the\footline}
               \vskip -.15truecm\line{\the\date\hfil}
              \else\line{\the\footline}\fi}
\paperheadline={
\iffrontpage\hfil
               \else
               \tenrm\hss $-$\ \folio\ $-$\hss\fi    }
\def\sectionfont{\bf}    }
%
%
%
%
\def\SIZE{\hsize=15.73truecm\vsize=23.11truecm}
\def\OFFSET{\voffset=0.0truecm\hoffset=0.truecm}
\message{DEFAULT FORMAT}
\def\papersize{\SIZE\OFFSET\skip\footins=\bigskipamount
\normaldisplayskip= 35pt plus 3pt minus 7pt}
\Pubnum={\rm NIKHEF--H/\the\pubnum }
\def\title#1{\vskip\frontpageskip\vskip .50truein
     \titlestyle{\seventeencp #1} \vskip\headskip\vskip\frontpageskip
     \vskip .2truein}
\def\author#1{\vskip .27truein\titlestyle{#1}\nobreak}
\def\andauthor{\vskip .27truein\centerline{and}\author}
\def\p@bblock{\begingroup \tabskip=\hsize minus \hsize
   \baselineskip=1.5\ht\strutbox \topspace+2\baselineskip
   \halign to\hsize{\strut ##\hfil\tabskip=0pt\crcr
  \the \Pubnum \cr hep-th/9410010 \cr}\endgroup}
\def\makefootline{\iffrontpage\vskip .27truein\line{\the\footline}
                 \vskip -.1truein
              \else\line{\the\footline}\fi}
\paperfootline={\iffrontpage\message{FOOTLINE}
\hfil\else\hfil\fi}

\def\abstract{\vskip2\frontpageskip\centerline{\twelvebf Abstract}
              \vskip\headskip }

\paperheadline={
\iffrontpage\hfil
               \else
               \twelverm\hss $-$\ \folio\ $-$\hss\fi}
%
%
\def\nup#1({\refbreak\ Nucl.\ Phys.\ $\underline {B#1}$\ (}
\def\plt#1({\refbreak\ Phys.\ Lett.\ $\underline  {#1}$\ (}
\def\cmp#1({\refbreak\ Commun.\ Math.\ Phys.\ $\underline  {#1}$\ (}
\def\prp#1({\refbreak\ Physics\ Reports\ $\underline  {#1}$\ (}
\def\prl#1({\refbreak\ Phys.\ Rev.\ Lett.\ $\underline  {#1}$\ (}
\def\prv#1({\refbreak\ Phys.\ Rev. $\underline  {D#1}$\ (}
\def\und#1({            $\underline  {#1}$\ (}
%
%

\def\rB{\hfil\penalty1000\hfilneg}
%
%
\hyphenation{sym-met-ric anti-sym-me-tric re-pa-ra-me-tri-za-tion
Lo-rentz-ian a-no-ma-ly di-men-sio-nal two-di-men-sio-nal}
%
%
%
%

\def\coeff#1#2{{\textstyle { #1 \over #2}}\displaystyle}
\def\boxit#1{\vbox{\hrule\hbox{\vrule\kern3pt
\vbox{\kern3pt#1\kern3pt}\kern3pt\vrule}\hrule}}
\message{ by V.K, W.L and A.S}
\catcode`@=12
\paperstyle

\def\CIZ{\rrr\CIZ{A.~Cappelli, C.~Itzykson and J.-B.~Zuber,
\nup280 (1987)  445; \hfill\break \cmp113 (1987) 1.}}
\def\FQS{\rrr\FQS{D.~Friedan, Z.~Qiu and S.~Shenker, \prl52  (1984) 1575.}}
\def\AhWa{\rrr\Ahwa{C.\ Ahn and M.\ Walton, \plt B223 (1989) 343.}}
\def\FGK {\rrr\FGK {G.~Felder, K.~Gawedzki and A.~Kupiainen,
\cmp 117 (1988) 127.}}
\def\KaPe{\rrr\KaPe{ V.G.~Kac and D.H.~Peterson, Adv.~Math.~53 (1984) 125.}}
\def\CoSua{\rrr\CoSua{F.~Bais and P.~Bouwknegt, \nup279 (1987) 561.}}
\def\CoSub{\rrr\CoSub{ A.N.~Schellekens and N.P.~Warner, \prv34 (1986) 3092.}}
\def\DijV{\rrr\DijV{R.~Dijkgraaf and E.~Verlinde, \rB
Nucl.~Phys.~(Proc.~Suppl.) \und{5B} (1988) 87.}}
\def\ScYe{\rrr\ScYe{A.N.~Schellekens and S.~Yankielowicz, \nup334 (1990) 67.}}
\def\BoNa{\rrr\BoNa{P.~Bouwknegt and W.~Nahm, \plt B184 (1987) 359.}}
\def\MoSb{\rrr\MoSb{G.~Moore   and N.~Seiberg, \nup313 (1988) 16.}}
\def\Scha{\rrr\Scha{A.N.~Schellekens, \plt B244 (1990) 255.}}
\def\ScYb{\rrr\ScYb{A.N.~Schellekens and S.~Yankielowicz,
\plt B227 (1989) 387.}}
\def\ScYa{\rrr\ScYa{A.N.~Schellekens and S.~Yankielowicz,
\nup 327 (1989) 673.}}
\def\Itz{\rrr\Itz{C.~Itzykson, Nucl.~Phys.~(Proc.~Suppl) \und{5B} (1988) 150.}}
\def\Ber{\rrr\Ber{D.~Bernard, \nup288 (1987) 628.}}
\def\ALZ {\rrr\ALZ{D.~Altschuler, J.~Lacki and P.~Zaugg, \plt B205 (1988)
281.}}
\def\Alig{\rrr\Alig{K.~Intriligator, \nup 332 (1990) 541.}}
\def\Fuch{\rrr\Fuch{J.~Fuchs, \cmp 136 (1991) 345.}}
\def\BeBe{\rrr\BeBe {B.~Gato-Rivera and A.N.~Schellekens, \nup 353 (1991)
519.}}
\def\BeBT{\rrr\BeBT {B.~Gato-Rivera and A.N.~Schellekens,
Commun.~Math.~Phys.~\und{145} (1992) 85.}}
\def\Ver {\rrr\Ver  {D.~Verstegen, \nup 346 (1990) 349; \cmp 137 (1991) 567.}}
\def\KaWa{\rrr\KaWa{V.G.~Kac and M.~Wakimoto, Adv.~Math.~\und{70} (1988) 156.}}
\def\Walt{\rrr\Walt{M.~Walton, \nup322 (1989) 775.}}
\def\ABI {\rrr\ABI {D.~Altschuler, M.~Bauer and C.~Itzykson,
\cmp132 (1990) 349.}}
\def\SchM{\rrr\SchM{A.N.~Schellekens, \cmp 153 (1993)  159.  }}
\def\SchK{\rrr\SchK{M.~Kreuzer and A.N.~Schellekens, \nup411 (1994) 97.}}
\def\FGSS{\rrr\FGSS {J.~Fuchs, B.~Gato-Rivera, A.N.~Schellekens and
C.~Schweigert, \plt B334 (1994) 113.}}
\def\dBG{\rrr\dBG{J.~de Boer and J.~Goeree, \cmp 139 (1991) 267.}}
\def\CoGa{\rrr\CoGa{A.~Coste and T.~Gannon, \plt B323 (1994) 316.}}
\def\Gan{\rrr\Gan{T.~Gannon, \cmp 161 (1994) 233.}}
\def\GanB{\rrr\GanB{T.~Gannon, {\it The Classification
of $SU(N)_k$ automorphism invariants}, \hfill\break hep-th/9212060.}}
\def\GanA{\rrr\GanA{T.~Gannon, \nup 396 (1993) 708.}}
\def\ScYg{\rrr\ScYg{A.N.~Schellekens and S.~Yankielowicz,
Int.~J.~Mod.~Phys.~\und{A5} (1990) 2903.}}
\def\Sal{\rrr\Sal{V.\ Pasquier and H.\ Saleur, \nup 330 (1990)  523.}}
\def\Warn{\rrr\Warn{N.~Warner, \cmp 190 (1990) 205.}}
\def\RoTe{\rrr\RoTe{P.~Roberts and H.~Terao, Int.~J.~Mod.~Phys.~\und{A7}
 (1992) 2207.}}
\def\NaSc{\rrr\NaSc{S.~Naculich and H.~Schnitzer, \nup 347 (1990) 687; \rB
                   \plt B244 (1990) 235.}}
\def\AbAr{\rrr\AbAr{M.~Abolhassani and F.~Ardalan,
Int.~J.~Mod.~Phys.~\und{A9} (1994) 2707.}}
\def\FSch{\rrr\FSch {J.~Fuchs and C.~Schweigert, Ann.~Phys.~\und{234} (1994)
102.}}
%
\newbox\hdbox%
\newcount\hdrows%
\newcount\multispancount%
\newcount\ncase%
\newcount\ncols
\newcount\nrows%
\newcount\nspan%
\newcount\ntemp%
\newdimen\hdsize%
\newdimen\newhdsize%
\newdimen\parasize%
\newdimen\spreadwidth%
\newdimen\thicksize%
\newdimen\thinsize%
\newdimen\tablewidth%
\newif\ifcentertables%
\newif\ifendsize%
\newif\iffirstrow%
\newif\iftableinfo%
\newtoks\dbt%
\newtoks\hdtks%
\newtoks\savetks%
\newtoks\tableLETtokens%
\newtoks\tabletokens%
\newtoks\widthspec%
%
%
\immediate\write15{%
CP SMSG GJMSINK TEXTABLE --> TABLE MACROS V. 851121 JOB = \jobname%
}%
%
%
\tableinfotrue%
\catcode`\@=11
%
%
\def\tstrut{\vrule height3.1ex depth1.2ex width0pt}%
\def\and{\char`\&}
\def\tablerule{\noalign{\hrule height\thinsize depth0pt}}%
\thicksize=1.5pt
\thinsize=0.6pt
\def\thickrule{\noalign{\hrule height\thicksize depth0pt}}%
\def\ctr#1{\hfil\ #1\hfil}%
%
%
%
%
\tablewidth=-\maxdimen%
\spreadwidth=-\maxdimen%
\def\tabskipglue{0pt plus 1fil minus 1fil}%
%
%
\centertablestrue%
%
%
%
%
\parasize=4in%
\gdef\ARGS{########}
\gdef\headerARGS{####}
\def\@mpersand{&}
{\catcode`\|=13
\gdef\letbarzero{\let|0}
\gdef\letbartab{\def|{&&}}%
\gdef\letvbbar{\let\vb|}%
}
{\catcode`\&=4
\def\ampskip{&\omit\hfil&}
\catcode`\&=13
\let&0
\xdef\letampskip{\def&{\ampskip}}%
\gdef\letnovbamp{\let\novb&\let\tab&}
}
\def\begintable{
   \begingroup%
   \catcode`\|=13\letbartab\letvbbar%
   \catcode`\&=13\letampskip\letnovbamp%
   \def\multispan##1{
      \omit \mscount##1%
      \multiply\mscount\tw@\advance\mscount\m@ne%
      \loop\ifnum\mscount>\@ne \sp@n\repeat%
   }
   \def\|{%
      &\omit\widevline&%
   }%
   \ruledtable
}
\long\def\ruledtable#1\endtable{%
%
%
%
   \offinterlineskip
   \tabskip 0pt
   \def\widevline{\vrule width\thicksize}
   \def\endrow{\@mpersand\omit\hfil\crnorm\@mpersand}%
   \def\crthick{\@mpersand\crnorm\thickrule\@mpersand}%
   \def\crthickneg##1{\@mpersand\crnorm\thickrule
          \noalign{{\skip0=##1\vskip-\skip0}}\@mpersand}%
   \def\crnorule{\@mpersand\crnorm\@mpersand}%
   \def\crnoruleneg##1{\@mpersand\crnorm
          \noalign{{\skip0=##1\vskip-\skip0}}\@mpersand}%
   \let\nr=\crnorule
   \def\endtable{\@mpersand\crnorm\thickrule}%
   \let\crnorm=\cr
%
%
   \edef\cr{\@mpersand\crnorm\tablerule\@mpersand}%
   \def\crneg##1{\@mpersand\crnorm\tablerule
          \noalign{{\skip0=##1\vskip-\skip0}}\@mpersand}%
   \let\ctneg=\crthickneg
   \let\nrneg=\crnoruleneg
   \the\tableLETtokens
%
%
   \tabletokens={&#1}
%
%
   \countROWS\tabletokens\into\nrows%
   \countCOLS\tabletokens\into\ncols%
%
%
   \advance\ncols by -1%
   \divide\ncols by 2%
   \advance\nrows by 1%
%
%
   \iftableinfo %
      \immediate\write16{[Nrows=\the\nrows, Ncols=\the\ncols]}%
   \fi%
%
%
   \ifcentertables
      \ifhmode \par\fi
      \line{
      \hss
   \else %
      \hbox{%
   \fi
      \vbox{%
         \makePREAMBLE{\the\ncols}
         \edef\next{\preamble}
         \let\preamble=\next
         \makeTABLE{\preamble}{\tabletokens}
      }
      \ifcentertables \hss}\else }\fi
   \endgroup
   \tablewidth=-\maxdimen
   \spreadwidth=-\maxdimen
}
\def\makeTABLE#1#2{
   {
   \let\ifmath0
   \let\header0
   \let\multispan0
%
%
   \ncase=0%
   \ifdim\tablewidth>-\maxdimen \ncase=1\fi%
   \ifdim\spreadwidth>-\maxdimen \ncase=2\fi%
   \relax
%
   \ifcase\ncase %
      \widthspec={}%
   \or %
      \widthspec=\expandafter{\expandafter t\expandafter o%
                 \the\tablewidth}%
   \else %
      \widthspec=\expandafter{\expandafter s\expandafter p\expandafter r%
                 \expandafter e\expandafter a\expandafter d%
                 \the\spreadwidth}%
   \fi %
   \xdef\next{
      \halign\the\widthspec{%
      #1
      \noalign{\hrule height\thicksize depth0pt}
      \the#2\endtable
%
      }
   }
   }
   \next
}
\def\makePREAMBLE#1{
   \ncols=#1
   \begingroup
   \let\ARGS=0
   \edef\xtp{\widevline\ARGS\tabskip\tabskipglue%
   &\ctr{\ARGS}\tstrut}
   \advance\ncols by -1
   \loop
      \ifnum\ncols>0 %
      \advance\ncols by -1%
      \edef\xtp{\xtp&\vrule width\thinsize\ARGS&\ctr{\ARGS}}%
   \repeat
   \xdef\preamble{\xtp&\widevline\ARGS\tabskip0pt%
   \crnorm}
   \endgroup
}
\def\countROWS#1\into#2{
   \let\countREGISTER=#2%
   \countREGISTER=0%
   \expandafter\ROWcount\the#1\endcount%
}%
\def\ROWcount{%
   \afterassignment\subROWcount\let\next= %
}%
\def\subROWcount{%
   \ifx\next\endcount %
      \let\next=\relax%
   \else%
      \ncase=0%
      \ifx\next\cr %
         \global\advance\countREGISTER by 1%
         \ncase=0%
      \fi%
      \ifx\next\endrow %
         \global\advance\countREGISTER by 1%
         \ncase=0%
      \fi%
      \ifx\next\crthick %
         \global\advance\countREGISTER by 1%
         \ncase=0%
      \fi%
      \ifx\next\crnorule %
         \global\advance\countREGISTER by 1%
         \ncase=0%
      \fi%
      \ifx\next\crthickneg %
         \global\advance\countREGISTER by 1%
         \ncase=0%
      \fi%
      \ifx\next\crnoruleneg %
         \global\advance\countREGISTER by 1%
         \ncase=0%
      \fi%
      \ifx\next\crneg %
         \global\advance\countREGISTER by 1%
         \ncase=0%
      \fi%
      \ifx\next\header %
         \ncase=1%
      \fi%
      \relax%
      \ifcase\ncase %
         \let\next\ROWcount%
      \or %
         \let\next\argROWskip%
      \else %
      \fi%
   \fi%
   \next%
}
\def\counthdROWS#1\into#2{%
\dvr{10}%
   \let\countREGISTER=#2%
   \countREGISTER=0%
\dvr{11}%
\dvr{13}%
   \expandafter\hdROWcount\the#1\endcount%
\dvr{12}%
}%
\def\hdROWcount{%
   \afterassignment\subhdROWcount\let\next= %
}%
\def\subhdROWcount{%
   \ifx\next\endcount %
      \let\next=\relax%
   \else%
      \ncase=0%
      \ifx\next\cr %
         \global\advance\countREGISTER by 1%
         \ncase=0%
      \fi%
      \ifx\next\endrow %
         \global\advance\countREGISTER by 1%
         \ncase=0%
      \fi%
      \ifx\next\crthick %
         \global\advance\countREGISTER by 1%
         \ncase=0%
      \fi%
      \ifx\next\crnorule %
         \global\advance\countREGISTER by 1%
         \ncase=0%
      \fi%
      \ifx\next\header %
         \ncase=1%
      \fi%
\relax%
      \ifcase\ncase %
         \let\next\hdROWcount%
      \or%
         \let\next\arghdROWskip%
      \else %
      \fi%
   \fi%
   \next%
}%
{\catcode`\|=13\letbartab
\gdef\countCOLS#1\into#2{%
   \let\countREGISTER=#2%
   \global\countREGISTER=0%
   \global\multispancount=0%
   \global\firstrowtrue
   \expandafter\COLcount\the#1\endcount%
   \global\advance\countREGISTER by 3%
   \global\advance\countREGISTER by -\multispancount
}%
\gdef\COLcount{%
   \afterassignment\subCOLcount\let\next= %
}%
{\catcode`\&=13%
\gdef\subCOLcount{%
   \ifx\next\endcount %
      \let\next=\relax%
   \else%
      \ncase=0%
      \iffirstrow
         \ifx\next& %
            \global\advance\countREGISTER by 2%
            \ncase=0%
         \fi%
         \ifx\next\span %
            \global\advance\countREGISTER by 1%
            \ncase=0%
         \fi%
         \ifx\next| %
            \global\advance\countREGISTER by 2%
            \ncase=0%
         \fi
         \ifx\next\|
            \global\advance\countREGISTER by 2%
            \ncase=0%
         \fi
         \ifx\next\multispan
            \ncase=1%
            \global\advance\multispancount by 1%
         \fi
         \ifx\next\header
            \ncase=2%
         \fi
         \ifx\next\cr       \global\firstrowfalse \fi
         \ifx\next\endrow   \global\firstrowfalse \fi
         \ifx\next\crthick  \global\firstrowfalse \fi
         \ifx\next\crnorule \global\firstrowfalse \fi
         \ifx\next\crnoruleneg \global\firstrowfalse \fi
         \ifx\next\crthickneg  \global\firstrowfalse \fi
         \ifx\next\crneg       \global\firstrowfalse \fi
      \fi
\relax
      \ifcase\ncase %
         \let\next\COLcount%
      \or %
         \let\next\spancount%
      \or %
         \let\next\argCOLskip%
      \else %
      \fi %
   \fi%
   \next%
}%
\gdef\argROWskip#1{%
   \let\next\ROWcount \next%
}
\gdef\arghdROWskip#1{%
   \let\next\ROWcount \next%
}
\gdef\argCOLskip#1{%
   \let\next\COLcount \next%
}
}
}
\def\spancount#1{
   \nspan=#1\multiply\nspan by 2\advance\nspan by -1%
   \global\advance \countREGISTER by \nspan
   \let\next\COLcount \next}%
\def\dvr#1{\relax}%
\def\header#1{%
\dvr{1}{\let\cr=\@mpersand%
\hdtks={#1}%
\counthdROWS\hdtks\into\hdrows%
\advance\hdrows by 1%
\ifnum\hdrows=0 \hdrows=1 \fi%
\dvr{5}\makehdPREAMBLE{\the\hdrows}%
\dvr{6}\getHDdimen{#1}%
{\parindent=0pt\hsize=\hdsize{\let\ifmath0%
\xdef\next{\valign{\headerpreamble #1\crnorm}}}\dvr{7}\next\dvr{8}%
}%
}\dvr{2}}
\def\makehdPREAMBLE#1{
\dvr{3}%
\hdrows=#1
{
\let\headerARGS=0%
\let\cr=\crnorm%
\edef\xtp{\vfil\hfil\hbox{\headerARGS}\hfil\vfil}%
\advance\hdrows by -1
\loop
\ifnum\hdrows>0%
\advance\hdrows by -1%
\edef\xtp{\xtp&\vfil\hfil\hbox{\headerARGS}\hfil\vfil}%
\repeat%
\xdef\headerpreamble{\xtp\crcr}%
}
\dvr{4}}
\def\getHDdimen#1{%
\hdsize=0pt%
\getsize#1\cr\end\cr%
}
\def\getsize#1\cr{%
\endsizefalse\savetks={#1}%
\expandafter\lookend\the\savetks\cr%
\relax \ifendsize \let\next\relax \else%
\setbox\hdbox=\hbox{#1}\newhdsize=1.0\wd\hdbox%
\ifdim\newhdsize>\hdsize \hdsize=\newhdsize \fi%
\let\next\getsize \fi%
\next%
}%
\def\lookend{\afterassignment\sublookend\let\looknext= }%
\def\sublookend{\relax%
\ifx\looknext\cr %
\let\looknext\relax \else %
   \relax
   \ifx\looknext\end \global\endsizetrue \fi%
   \let\looknext=\lookend%
    \fi \looknext%
}%
%
%
\def\tablelet#1{%
   \tableLETtokens=\expandafter{\the\tableLETtokens #1}%
}%
\catcode`\@=12
%

\paperstyle
\def\DC{g}
\def\GCD{{\rm GCD}}
\def\Zbf{{\bf Z}}
\def\mod{{\rm ~mod~}}

\def\half{\coeff12}

\def\X{{\cal X}}
\def\ie{i.e.}

\pubnum={{94--31}}
\date{October 1994}
\pubtype{CRAP}
\titlepage
\message{TITLE}
\title{\fourteenbf Galois Modular Invariants of WZW Models}
\author{~~J. Fuchs $^{\times}$}
\andauthor{A. N. Schellekens}
\andauthor{C. Schweigert}
\bigskip
\line{\hfil \tenpoint\it NIKHEF-H / FOM, Postbus 41882, 1009$\,$DB Amsterdam,
The Netherlands  \hfil}
\abstract
\noindent The set of modular invariants that can be obtained from
Galois transformations is investigated systematically for
WZW models. It is shown that a large subset of Galois modular invariants
coincides with simple current invariants. For algebras of type $B$ and $D$
infinite series of previously unknown exceptional automorphism invariants
are found. \vfill
\noindent------------------------ \hfill\break
$^{\times}\,$ Heisenberg fellow \bigskip

\endpage

\chapter{Introduction}
The problem of finding all modular invariant partition functions of
rational conformal field theories (RCFT's) remains to a large extent
unsolved. This problem is part of the
programme of classifying all rational conformal field theories, which in
turn is part of the even more ambitious programme of classifying all
string theories.

The aim is to find a matrix $P$ that commutes with the generators
$S$ and $T$ of the modular group, and that furthermore is integer-valued,
non-negative and has $P_{00}=1$, where $0$ represents the identity
primary field. The partition function of the theory has then the form
$ \sum_{i j}\X_i P_{ij} \bar{\X}_j^* $, where $\X_i$ are the characters of
the left chiral algebra and $\bar \X_j$ those of the right one (the left and
right algebras need not necessarily coincide).

At present the classification is complete only for
the simplest RCFT's, whose chiral algebra consists only of the
Virasoro algebra \FQS,\CIZ. The next simplest case is that of
WZW models, whose chiral algebra has in addition to the
Virasoro algebra further currents of spin 1. In general such a theory can
be `heterotic' (\ie\ it  may have different left and right Kac-Moody algebras)
and both the left and right chiral algebra may have more than one
affine factor, but even in the simplest case -- equal left and right
simple affine algebras -- the classification
is complete at arbitrary level only for the cases $A_1$ \CIZ\ and $A_2$
\Gan.  Several other partial classification results have been presented,
see for example \multref\Itz{\GanA\GanB}.

Although there is no complete classification, many methods are known for
finding at least a substantial number of solutions, for example simple
currents \ScYa\ (see also \multref\Ber{\ALZ\AhWa\Alig\FGK}),
conformal embeddings \BoNa, rank-level duality
\multref\KaWa{\Walt\ABI\NaSc\Ver\FSch}, supersymmetric index arguments
\Warn, selfdual lattice methods \RoTe, orbifold constructions using
discrete subgroups of Lie groups \AbAr, and
the elliptic genus \SchM. In a previous paper \FGSS, we introduced an
additional method based on Galois symmetry of the matrix $S$ of a RCFT,
a symmetry that was discovered by de Boer and Goeree \dBG\ and further
investigated by Coste and Gannon \CoGa. This work will be reviewed briefly in
the next chapter.  The main purpose of this paper is to study in more
detail the application of this new method to WZW models.

Galois symmetry organizes the fields of a CFT into orbits, and along these
orbits the matrix elements of $S$ are algebraically conjugate numbers. Based on
this knowledge we are able to write down a number of integer-valued matrices
$P$ that commute with $S$, but do not necessarily commute with $T$ and are
not necessarily positive. These matrices span what we call the
`Galois-commutant' of $S$. This commutant can be constructed
in a straightforward manner from the Galois orbits, which in turn
can be obtained by scaling vectors in weight space by certain integers,
and mapping them back into the fundamental affine Weyl chamber (for
a more precise formulation we refer to chapter \ChapGal\ and the appendix).
This
is a simple algorithm that can be carried out easily with the help of
a computer. The time required for this computation increases linearly
with the number of primary fields, and for each primary
the number of calculational steps is bounded from above by
the order of the Weyl group.  This should be compared with the computation of
the modular matrix $S$, which grows quadratically with the number of primaries,
and which requires a sum over the full Weyl group (although several
shortcuts exist, for example simple currents and of course Galois symmetry).

Our second task is then to find the positive $T$-invariants within
the Galois commutant. In some cases this can be done analytically.
This class, which contains only simple current invariants, is
discussed in chapter \ChapInf.  In general however one has to
solve a set of equations for a number of integer coefficients. The
number of unknowns can grow rather rapidly with increasing level of the
underlying affine Kac-Moody algebra
-- Galois symmetry is a huge and very powerful symmetry -- which
is another limitation on the scope of our investigations.

In practice we have considered algebras with rank $\leq 8$ and up to
2500 primary fields, but this range was extended when there was reason to
expect
something interesting. Although a lot of exploratory work has already
been done on the classification of modular invariants, only fairly
recently new invariants were found \SchM\ for $E_6$
and $E_7$ at rather low levels (namely 4 and 3), showing that there are
still chances for finding something new. Indeed, we did find new invariants,
namely an infinite series of exceptional automorphism invariants for
algebras of type $B$ at level 2, starting at rank 7, as well as for
algebras of type $D$ at level 2.
In addition we find for the same algebras some clearly unphysical
extensions by spin-1 currents.  This is explained in chapter \ChapNew.
Other exceptional invariants that can be explained in terms of Galois
symmetry are presented in chapter \ChapPure.

We have also considered the possibility of combining Galois orbits
with simple current orbits. In chapter \ChapExt\ we discuss two ways of
doing that, one of which is to apply Galois symmetry to simple
current extensions of the chiral algebra.

To conclude this introduction we fix some notations.
If $P_{0i}=P_{i0}=0$ for all $i \neq 0$, the matrix $P$ defines a permutation
of the fields in the theory that leaves the fusion rules invariant. We will
refer to this as an {\it automorphism invariant}.
Under multiplication such matrices form a group which  is a subgroup of
the group of {\it fusion rule automorphisms}.  These are all permutations of
the fields that leave the fusion rules
invariant, but which do not necessarily  commute with $S$ or $T$.
Finally there is a third group of automorphisms we will encounter, namely that
of {\it Galois automorphisms}. They act as a permutation combined with
sign flips, and may act non-trivially on the identity.  It is important not to
confuse these three kinds of automorphisms.

If a matrix $P$ does not have the form of an automorphism invariant,
and if the partition function is a sum of squares of linear
combinations of characters, we will
refer to it as a ({\it chiral algebra}) {\it extension}.  If it is not a sum of
squares it can be viewed as an automorphism invariant of an extended
algebra \multref\MoSb{\DijV}\ (at least if an associated CFT exists).

A matrix $P$ corresponding to a chiral algebra extension may contain
squared terms  appearing with a multiplicity higher than 1. Such terms
will be referred to as `fixed points', a terminology which up to now was
appropriate only for extensions by simple currents. Galois automorphisms
provide us with a second rationale for using this name. Usually such fixed
points correspond to more than one field in the extended CFT, and
they have to be `resolved'.  The procedure for doing this is available
only in some cases, and then only for $S$, $T$, the fusion rules and in
a few cases also the characters \ScYe.

\chapter{Galois Symmetry in Conformal Field Theory}

As is well known, a rational \cft\ gives rise to a \findim\ unitary
representation of
$SL_2(\zett)$, the double cover of the modular group. Namely, given a
rational fusion ring with generators $\phi_i,\ i\in I$ ($I$ some finite
index set), and relations
  $$  \phi_i \times \phi_j = \sumik {\cal N}_{ij}^{\ k} \phi_k \,, $$
there is a unitary and symmetric matrix $S$ that diagonalizes the
fusion matrices, i.e.\ the matrices ${\cal N}_i$ with entries
$({\cal N}_i)^k_j:= {\cal N}_{ij}^{\ k}$. $S$ and the matrix $T$ with entries
$T_{ij}:= \eE^{2\pi \ii (h_i - c/24)}\delta_{ij}$
(with $h_i$ the conformal weights and $c$ the conformal central charge),
generate a representation of $SL_2(\zett)$. In particular, $S^2=C=(ST)^3$
where $C$, the charge conjugation matrix, is a permutation of order
two, which we write as $C_{ij}^{}=\delta^{}_{i,j^+}$.
By the Verlinde formula
   $$  {\cal N}_{ij}^{\ k} = \sumil\, {S^{}_{i\ell}S^{}_{j\ell}S^*_{k\ell}
   \over S_{0\ell}^{}} \,,  $$
the eigenvalues of the fusion matrices ${\cal N}_i$ are the
generalized {\it quantum dimensions\/}\hfill\break $S_{ij}/S_{0j}$;
the label $0$ refers to the identity primary field; it satisfies
$0=0^+$ and corresponds to the unit of the fusion ring. The quantum dimensions
realize all inequivalent irreducible representations of the fusion ring
(which are \onedim), i.e.\ we have
   $$ {S_{i\ell} \over S_{0\ell}}\, {S_{j\ell} \over S_{0\ell}} =
  \sumik\, {\cal N}_{ij}^{\ k}\, {S_{k\ell} \over S_{0\ell}}  \eqn\odim $$
for all $\ell\in I$.

The \qdim s are the roots of the characteristic polynomial
  $$  \det(\lambda \one - {\cal N}_i) \,.$$
This polynomial has integral coefficients
and is normalized, \ie\ its leading coefficient is equal to 1. As a
consequence, the \qdim s are algebraically integer numbers in some
algebraic number field $L$ over the rational numbers \rationals. The
extension $L/\rationals$ is normal \dBG;
since the field $\rationals$ has characteristic zero, this implies that it
is a {\em Galois} extension; its {\it\gagr}, denoted by $\,\gallq$,
is abelian. Invoking the theorem of Kronecker and Weber, this shows
that $L$ is contained in some cyclotomic field $\rationals(\zeta_n)$,
where $\zeta_n$ is a primitive $n$th root of unity.

By applying an element $\sigmaL \in \Gal(L/\rationals)$ on equation \odim\
it follows that the numbers $\sigmaL(S_{ij}/S_{0j})$,
$i\in I$, again realize a \onedim\ representation of the fusion ring.
As the (generalized) quantum dimensions exhaust all inequivalent \onedim\
representations of the fusion ring, it follows that there exists
some permutation $\sigmA$ of the index set $I$ such that
  $$    \sigmaL({S_{ij} \over S_{0j}})
  = {S_{i, \sigmA j } \over S_{0, \sigmA j }} \,.$$

To obtain an action of a Galois group on the entries $S_{ij}$ of the \smat,
rather than just on the \qdim s, one has to
consider also the field $M$ which is the extension
of \rationals\ that is generated by all \smat\ elements.
$M$ extends $L$ as well, and
the extension $M/\rationals$ is again normal and has abelian \gagr.
It follows that $\galml$ is a normal subgroup of $\galmq$ and
that the sequence $ 0 \to \galml \to \galmq \to \gallq \to 0 \,,$
where the second map is the canonical inclusion and the third one
the restriction map, is exact. Therefore
  $$  \Gal(L/\rationals) \cong \Gal(M/\rationals) \,/\, \Gal(M/L) \,.  $$
In particular, upon restriction from $M$ to $L$
any $\sigmaM \in \Gal(M/\rationals)$
maps $L$ onto itself and coincides with some element
$\sigmaL \in \Gal(L/\rationals)$, and conversely, any
$ \sigmaL \in \Gal(L/\rationals)$ can be obtained this way.
Correspondingly, as in \FGSS\ we will frequently use the abbreviation
$\sigma$ for both $\sigmaM$ and its restriction $\sigmaL$.

For any $\sigma\in \gallq$ there exist \CoGa\
signs $\epsilon_\sigma(i) \in\{\pm1\}$ such that the relation
  $$  \sigma(S_{ij}) = \epss(i)\cdot S_{\sigmA i ,j}  \eqn\ssij $$
holds for all $i,j\in I$. While the order $N$ of the Galois group
element $\sigma$ and the order $\dot N$ of
the permutation $\sigmA$ of the labels that is induced by $\sigma$
need not necessarily coincide, only an extra factor of 2 can appear,
and the elements with $N=2\dot N$ turn out to be quite uninteresting \FGSS.

Let us now describe a few elementary facts about Galois theory of
cyclotomic fields. Denote by $\zett_n^*$ the multiplicative group of all
elements of $\zett_n\equiv\zett/n\zett$ that are coprime with $M$. Note
that precisely these
elements have an inverse with respect to multiplication. (For example,
the group $(\zett_{10}^*,\cdot)\cong(\{\pm1,\pm3\},\cdot\;{\rm mod}\,10)$
is isomorphic to the additive group $(\zett_4,+)$.)
The number $\varphi(n)$ of elements of $\zett_n^*$ is given by
Euler's $\varphi$ function, which can be computed as
follows. If $n=\prod_i p_i^{n_i}$ is a decomposition of $n$ into distinct
primes $p_i$, then one has
  $$  \varphi(n) = \varphi( \prod_i p_i^{n_i}) =
  \prod_i\varphi( p_i^{n_i}) =  \prod_i p_i^{n_i-1} (p_i -1) \,. $$

The Galois automorphisms (relative to
\rationals) of the cyclotomic field $\qzn$ in which
$\gallq$ is contained are in one-to-one correspondence with the elements
$\ell\in\zett_n^*$. The automorphism associated to each such $\ell$
simply acts as
  $$   \sigmal: \quad  \zeta_n \mapsto (\zeta_n)^\ell \,. $$
This implies in particular that $\ell=-1$ corresponds to
complex conjugation. Thus if the fusion ring is self-conjugate
in the sense that $i^+=i$ for all $i\in I$, so that the \smat\ is
real, then the automorphism $\sigma_{(-1)}$ acts
trivially. In this case the relevant field $L$ is already
contained in the maximal real subfield $\rationals(\zeta_n+\zeta_n^{-1})$
of the cyclotomic field $\qzn$, which is the field that is fixed
under complex conjugation.

In the special case where the fusion ring describes the fusion rules of a
\wzwt\ based on an affine Lie algebra g at level $k$, the Galois group
is a subgroup of $\zett_{M(k+g)}^*$, where $g$ is the dual Coxeter number
of the horizontal subalgebra of g and $M$ is the smallest positive integer
for which the numbers $MG_{ij}$, with $G_{ij}$ the entries of the metric on the
weight space of the horizontal subalgebra, are all integral.
A Galois transformation labelled by $\ell\in\zett_{M(k+g)}^*$ then induces
the permutation
  $$   \sigmAl(\Lambda)=\wh(\ell\cdot(\Lambda+\rho)-\rho) \,, \eqn\wzw $$
where $\Lambda$, the horizontal part of an integrable highest weight of
g at level $k$, labels the primary fields, $\rho$ is
the Weyl vector of the horizontal subalgebra, and where
$\wh$ is the horizontal projection of a suitable affine Weyl transformation.
The sign $\epsilon_{\sigmal}^{}$ is just given by the sign
of the Weyl transformation $\wh$, up to an overall sign $\eta$ that only
depends
on $\sigmal$, but not on $\Lambda$.  (For more details, see the appendix.)

\subsection{Fusion rule automorphisms} \Mskip

As has been shown in \FGSS, the properties of Galois transformations can be
employed to construct automorphisms of the fusion rules as well as
$S$-invariants. Consider first the case where
the permutation $\sigmA$ induced by the
Galois group element $\sigma$ leaves the identity fixed,
  $$   \sigmA\,0 =0 \,; $$
then $\sigmA$ is an automorphism of the fusion rules,
and the sign $\epss(i)$ is the same for all $i\in I$,
  $$  \epsilon_\sigma(i) = \epss(0) =: \epss  = {\rm const}  \,. $$
The presence of such automorphisms of the fusion rules can be
understood as follows. The `main' quantum dimensions
  $$ {S_{i0} \over S_{00}} $$
all lie in a real field $L_{(0)}$ that is contained in the field $L$
generated by all (generalized) quantum dimensions ${S_{ij} }/{ S_{0j}}$.
The elements of the group $\Gal(L/L_{(0)})$ leave the main quantum dimensions
invariant, and hence the associated permutations $\sigmA$ are fusion rule
automorphisms.

\subsection{$S$-invariants} \Mskip

To deduce $S$-invariants from these considerations it is convenient
to act on \ssij\ with $\sigma^{-1}$, and permuting the second label
of $S$ on the right hand side. Then we obtain the relation
  $$  S_{ij} =  \epsilon_\sigma(i) \epsilon_{\sigma^{-1}}(j) \,
  S_{\sigmA i , \sigmA^{-1} j } \,. \eqn\GaloisSymmetry $$
Now for any Galois transformation $\sigma$ we define the orthogonal matrix
$$ (\Pi_{\sigma})_{ij} := \vareps_{\sigma}(i)\, \delta_{j,\sigmA i}
            = \epsilon_{\sigma^{-1}}(j)\,\delta_{i, \sigmA^{-1} j} \ , $$
where in the second equality we used the relation
 $$\epsilon_{\sigma}(\sigmA^{-1}(i)) = \epsilon_{\sigma^{-1}}(i)$$
which is obtained from the identity $\,\sigma\sigma^{-1}S_{ij}=S_{ij}$
when acting twice on the first label of $S$.
These orthogonal matrices can easily be shown to satisfy the identities
$$ (\Pi_{\sigma})^{-1} = \Pi_{\sigma^{-1}}=(\Pi_{\sigma})^T \ , $$
and they implement the Galois transformations \ssij\ in the following way:
$$ \sigma S = \Pi_\sigma \cdot S = S \cdot \Pi_\sigma^{-1} \ .   $$
Now we can write \GaloisSymmetry\ in matrix notation as
(omitting the subscript $\sigma$ of $\Pi_\sigma$)
$$S=\Pi S \Pi\ , \eqn\PiSPi $$
or $\Pi^{-1} S = S \Pi $. Obviously the same identity
holds with $\Pi$ replaced by its inverse, and by adding these two
relations we see that the matrix $\Pi + \Pi^{-1} = \Pi + \Pi^T$
commutes with $S$. If $\Pi$ is equal to its own inverse one can take
half this matrix, \ie\ $\Pi$ itself.

The full Galois commutant is obtained by considering all sums and products
of these matrices. Because the matrices $\Pi$ form an abelian group (isomorphic
to the Galois group $\gallq$) it is easy to see that the product of any two
matrices of the form $\Pi + \Pi^{-1}$ is a linear combination
of such matrices with integral coefficients. Hence the most general
integer-valued $S$-invariant that can be obtained in this way is
$$ P=\sum_{(\sigma,\sigma^{-1}) \in G} f_{\sigma}
 (\Pi_{\sigma} + \Pi^{-1}_{\sigma} ) \ , \eqn\modinv$$
where the sum is over all elements of the Galois group $G$ modulo inversion,
and
$f_{\sigma} \in \Zbf$. This result was obtained before in \FGSS\ in a slightly
different formulation.

Note that this derivation of $S$-invariants goes through for any matrix
$\Pi$ that satisfies \PiSPi, even if it did not originate from Galois symmetry.
If such a new matrix $\Pi$ commutes with all matrices $\Pi_G$ that
represent Galois symmetries, one may extend the Galois group $G$ to
a larger group $\tilde G \supset G$  by including
all matrices $\Pi\cdot\Pi_G$. The most general $S$-invariant related to
$\tilde G$ is then obtained by extending the sum in \modinv\ to $\tilde G$.

As was observed in \CoGa, Galois symmetry implies a relation that any
modular invariant $P$, irrespective of whether it is itself a Galois invariant,
should satisfy. Indeed, using $\sigma P = P$ and $\sigma S^{-1}=
(\sigma S)^{-1}$, one derives
$P=\sigma P = \sigma (SPS^{-1})= \Pi_{\sigma} P \Pi^{-1}_{\sigma}$,
\ie\ $P$ commutes with $\Pi$. If $P$ is an automorphism of order 2, then we
have in addition the relation $S=PSP$, and hence $P$ is a `Galois-like'
automorphism that can be used to extend the Galois group as described
above. If $P$ is an automorphism of higher order or corresponds to
an extension of the chiral algebra, then it has
different commutation properties with $S$, and it cannot be used to
extend the Galois group, but one can still enlarge the commutant by
multiplying all matrices \modinv\ with the new invariant $P$ and its
higher powers. In this case the full commutant is considerably harder
to describe, however.

It must be noted that even if the matrix \modinv\ contains
negative entries, or does not commute with $T$,
it can still be relevant for the construction of physical modular invariants,
because the prescription may be combined with other procedures in such
a manner that the unwanted contributions cancel out.
For example one may use simple currents to extend the chiral algebra
before employing the Galois transformation. This will be discussed in
chapter \ChapExt.

\chapter{Infinite Series of Invariants}

\noindent In this chapter we will discuss an infinite class
of WZW modular invariants that can be obtained both by
a Galois scaling as well as by means of simple currents. Both Galois
transformations and simple currents organize the fields of a CFT into
orbits. In general, the respective orbits are not identical. In the special
case of WZW models which we focus on in this paper, these orbits are
in fact never identical, except for a few theories with too few
primary fields to make the difference noticeable. However, since the
orbits are used in quite different ways to derive modular invariants, it
can nevertheless happen that these invariants are the same.

The Galois scalings we consider are motivated by the following argument.
As already mentioned, Galois automorphisms of the fusion rules
arise if the field $L_{(0)}$ is strictly smaller than the field $L$.
In the case of WZW theories
$L$ is contained in the cyclotomic field $\rationals(\zeta_{M(k+g)})$
where $M$ is the denominator of the metric on weight space,
while the quantum Weyl formula \Sal\
  $$   {S_{a,\rho}\over S_{\rho,\rho}} = \prod_{\alpha>0} {
  \sin[\pi\,a\cdot\alpha^\vee/(k+g)] \over
  \sin[\pi\,\rho\cdot\alpha^\vee/(k+g)]} $$
shows that $L_{(0)}$ is already contained in $\rationals(\zeta_{2(k+g)})$.
Now as any element of $\gallq$ can be described by at least one element of
$\Gal(\rationals(\zeta_{M(k+g)})/\rationals)$, we do not loose
anything by working with the latter Galois group. Any Galois automorphism of
the
fusion rules can now be described by at least one element of
$\Gal(\rationals(\zeta_{M(k+g)})/L_{(0)})$. Unfortunately, $L_{(0)}$ is not
explicitly known in practice; therefore we would like to replace $L_{(0)}$
by the field $\rationals(\zeta_{2(k+g)})$ in which it is contained.
However, $M$ is not always even, and hence we consider instead of
$\rationals(\zeta_{2(k+g)})$ the smaller field $\rationals(\zeta_{k+g})$
and the corresponding Galois
group $\Gal(\rationals(\zeta_{M(k+g))}/\rationals(\zeta_{k+g}))$. The
elements of this group are precisely covered by scalings by a factor
$m(k+g)+1$. This way
we recover at least part of the automorphisms, but due to the
difference between $\rationals(\zeta_{2(k+g)})$ and $\rationals(\zeta_{k+g})$,
generically some of these scalings do not describe automorphisms, but
rather correspond to an extension of the chiral algebra.

Consider now the Kac-Peterson \KaPe\ formula
  $$ S_{ab} = {\cal N} \sum_w  {\varepsilon(w)}\,
  \exp[-2\pi\ii\,{w(a)\cdot b \over k+ \DC}] \  \eqn\moms $$
for the modular matrix $S$. Here ${\cal N}$ is a normalization factor which
follows by the unitarity of $S$ and is irrelevant for our purposes, and
the summation is over the Weyl group of the horizontal subalgebra of the
relevant affine Lie algebra; $a$ and $b$ are integrable
weights, shifted by adding the Weyl vector $\rho$.
In the following we will denote such shifted weights by roman characters
$a,b,\ldots\,$, while for the Lie algebra weights $a-\rho,\,b-\rho,\ldots$
we will use greek characters.

The scaling by a factor $\ell=m(k+g)+1$ is an allowed Galois
scaling if the following condition is fulfilled
(note that $m$ is defined modulo $M$):
  $$ ({\rm a}) \quad m(k+g)+1\ {\rm is\ prime\ relative\ to}\ M (k+g) \,.
  \hbox{~~~~~} $$
We will return to this condition later.
(Let us mention that even if condition (a) is not met, the scaling by $\ell$
can still be used to define an $S$-invariant. We will describe the
implications of such `quasi-Galois' scalings elsewhere.)

Under such a scaling  one has
$$ \eqalign{ S_{ab} \mapsto \sigma S_{ab}   & ={\cal N} \sum_w \varepsilon(w)\,
 \exp[-2\pi\ii\,{w(a)\cdot b \over k+ \DC} (m(k+g) + 1)] \cr
                & = \eE^{-2\pi\ii m a\cdot b } S_{ab}  \  ,\cr } \eqn\Scaling$$
where the last equality holds if $mw(a) \cdot b = m a\cdot b \mod1 $ for all
Weyl group elements $w$. To analyze when this condition is fulfilled, first
note that any Weyl transformation can be written as a product of
reflections with respect to the planes orthogonal to the simple roots.
For a Weyl reflection  $r_i$ with respect to a simple root
$\alpha_i$ ($i\in\{1,2,\ldots,{\rm rank}\}$) one has in general
  $$\eqalign{ r_{i}(a)\cdot b &= a\cdot b - ({ 
  2\over 
  {\alpha_i \cdot \alpha_i}}) \ \alpha_i\cdot a \  \alpha_i \cdot b \cr
&=a\cdot b - {1\over2} \,\alpha_i \cdot \alpha_i \,\,
   a_i b_i  \ , \cr } \eqn\SLcond$$
where $a_i$ and $b_i$ are Dynkin labels. Thus $r_{i}(a)\cdot b $ equals
$a\cdot b$ modulo integers if and only if all simple roots have norm 2
(which is for all algebras our normalization of the longest root), \ie\ iff the
algebra is simply laced. However, the derivation depends on this relation
with an extra factor
$m$. This yields one more non-trivial solution, namely $m=2$ for $B_n$,
$n$ odd. Note that for $B_n$ with $n$ even, one has $M=2$ so that the only
allowed scaling, $m=2$,  yields
a trivial solution. This is also true for all other non-simply laced algebras.

As is easily checked,
the quantity $a\cdot b \mod1$ is closely related to the product of the simple
current charges; we find:
$$ \eqalign{ A_n: ~~~~~~~~~~~~~~~~&a\cdot b = - (n+1) \Q(a)\Q(b) \cr
B_n:  ~~~~~~~~~~~~~~~~&2a\cdot b = 2n \Q(a) \Q(b) \cr
D_n\ (n {\rm~odd}):~~~~~&a\cdot b = 4 n  \Q(a)\Q(b) \cr
D_n\ (n {\rm~even}):~~~~&a\cdot b = 2  \Q_s(a)\Q_s(b)
        +2 \Q_c(a)\Q_c(a)  + (n-2) \Q_v(a)\Q_v(b) \cr
E_6: ~~~~~~~~~~~~~~~~&a\cdot b = 3 \Q(a)\Q(b) \cr
E_7:  ~~~~~~~~~~~~~~~~&a\cdot b = 2 \Q(a)\Q(b) \ . \cr }\eqn\ChargeRelations $$
Here  $\Q(a)$ is the monodromy charge  with respect to the simple current $J$
of a WZW representation with highest weight $a$ (which is at level $k+\DC$).
This should not be confused with the simple
current charge of the field labelled by $a$, which we denote
by $Q(a)$. The relation between these two quantities is
$$    Q(a)= \Q(a-\rho) = \Q(a) - \Q(\rho)  \ , \eqn\Qdiff $$
since the field labelled by $a$ has highest weight $a-\rho$ (which is at
level $k$). The charge $\Q$ (as well as $Q$) depends only on the
conjugacy class of the weight. The \wzwt\ with algebra $D_n, n$ even, has a
center $\Zbf_2 \times \Zbf_2$
and simple currents $J_s, J_v$ and $J_c=J_v \times J_s$. It has thus two
independent charges, for which one may take $Q_v$ and $Q_s$.

If $\rho$ is on the root lattice, then $\Q(\rho) = 0$ and
the shift  in \Qdiff\ is irrelevant, \ie\  $\Q=Q \mod 1$.
In general, either $\rho$ is a vector
on the root lattice, or it is a weight with the property that $2\rho$ is on
the root lattice. In the cases of interest here, $\rho$ is on the root
lattice for $A_n$, $n$ even, $D_n$ with $n=0 \mod 4 $ or $1 \mod 4$,
and for $E_6$. In all other cases
 $Q=\Q + \half \mod 1$ (if the algebra is $D_n, n = 2 \mod 4$, the charges
affected by this shift are $Q_s$ and $Q_c$).

Note that the left hand sides of the relations \ChargeRelations\ are
always of the form $l N \Q(a) \Q(b)$ or a sum of such terms, where $N$
is the order of the simple current and $l$ is an integer. The relation
for $B_n$ has an essential factor of 2 in the left hand side. Since
the relations are defined modulo integers we cannot simply divide this factor
out. The most convenient way to deal with it is to rewrite $m$ in this case
as $m=2\tilde m$ (we have already seen above that $m$ has
to be even for $B_n$).  After substituting
\ChargeRelations\ into \Scaling\ we get generically
$$ \sigma S_{ab} = \eE^{-2\pi\ii lm N \Q(a)\Q(b)} S_{ab} \ .\eqn\GaloisCharge$$
This formula holds for $B_n$ if one replaces $m$ by
$\tilde m$, and for $D_n, n$ even, if one replaces the
exponent by the appropriate sum, as in
\ChargeRelations. We will postpone the discussion of the latter case until
later, and consider for the moment only theories with a center $\Zbf_N$.

Now we wish to make use of the simple current relation
   $$ S_{J^na, b} = \eE^{2\pi\ii n Q(b)} S_{ab}\  .$$
This is simplest if we can replace $\Q$ by $Q$, and this
is the case we consider first. This replacement is allowed if $\rho$ is
on the root lattice, but this is not a necessary condition because of
the extra factor $lmN$. Suppose $\Q=Q+\half$. Then we see from the
foregoing that $N$ is even and $l$ odd.  Replacing $\Q$ by
$Q$ in the exponent of \GaloisCharge\
yields the extra terms $$\half lmNQ(a) + \half lmN Q(b) + \coeff{1}{4}
lmN \ , \eqn\ExtraTerms  $$ which should be an integer. Now
$N\Q(a)$ (or $N\Q(b)$) is an integer, which as a function of $a$ (or $b$)
takes all values modulo $N$.
Hence each of the three terms must separately be an integer. The first
two terms are integers if and only if $m$ is even. Then the last one
is an integer as well, since $N$ is even. Thus the condition that
$\Q$ can be replaced by $Q$ is equivalent to
  $$ ({\rm b}) \quad m\rho\ \hbox{is an element of the root lattice.
  ~~~~~~~~~~} $$
We remind the reader that for $B_n$ this is valid with $m$ replaced by
$\tilde m = \half m$. Hence condition (b) is in fact not satisfied for $B_n$
for
any non-trivial value of $m$. In all remaining algebras $M$ (the denominator
of the inverse symmetrized Cartan matrix) is equal to $N$.

If conditions (a) and (b) hold we can derive
$$ \sigma S_{ab} = S_{J^{-mlN Q(a)} a, b}
                            = S_{a,J^{-mlN Q(b)}b} \ . \eqn\GaloisSimple $$
On the other hand according to \ssij\ Galois invariance implies
$$ \sigma S_{ab} = \vareps_{\sigma}(a) S_{\dot{\sigma} a, b}=
  \vareps_{\sigma}(b) S_{a, \dot{\sigma} b}  \ . \eqn\GaloisEps $$
Furthermore if $m\rho$ is an element of the root lattice, it is easy to see
that the scale transformation fixes the identity field: the identity is
labelled by $\rho$, and transforms into
$\rho'=\rho + m(k+g)\rho$. The second term is a Weyl translation if
$m\rho$ is on the root lattice. In these cases $\rho'$ is mapped
to $\rho$ by the transformations described in the appendix, which implies
that the identity primary field is fixed. Then, according to \FGSS, it
follows that $\vareps \equiv 1$, and hence we find
$$ S_{J^{-mlNQ(a)}a,b} = S_{\dot{\sigma}a,b} \ , $$
or
$$ S_{a,b}=S_{\tau a,b} \ ,$$
where $\tau a = J^{mlNQ(a)}\dot{\sigma}a $. Then unitarity of $S$ implies
$ \delta_{a,\tau a}=\sum_b S_{\tau a,b} S_{ba}^* = \sum_b S_{ab} S_{ba}^* = 1,$
so that $a = \tau a$, and hence $\dot{\sigma} a = J^{-mlNQ(a)}a$.

As described in chapter \ChapGal, any Galois transform that fixes
the identity generates
an automorphism of the fusion rules, and in this case we see that it
connects fields on the same simple current orbit. It is a positive
$S$-invariant, but so far it was not required to respect
$T$-invariance. Thus the last condition we will now impose is
$$ ({\rm c})\quad  T\hbox{-invariance$\,.$~~~~~~~~~~~~~~~~~~~~~~~~~~~~} $$
In general for simple currents of order $N$ one has
$$ h(J^n a)=h(a)+h(J^n) -nQ(a) \mod 1 $$
and
$$ h(J^n)={rn(N-n) \over 2N} \mod 1 \ , $$
where $r$ is the monodromy parameter, which is equal to $k$ for $A_n$
at level $k$,
to $3nk\mod 8$ for $D_n$, $n$ odd, to $2k$ for $E_6$, and to $3k$ for $E_7$.
Condition (c) amounts to the requirement that the difference
$h(J^{-mlNQ(a)}a)-h(a)$ of conformal weights be an integer. We have
$$\eqalign{
  h(J^{-mlNQ(a)} a )&=h(a) +h(J^{-mlNQ(a)}) + mlNQ(a) Q(a)\cr
  & = h(a) - {r\over2 } mlNQ(a)  - ({r\over 2}  (ml)^2 - ml ) NQ(a)Q(a)\ .\cr }
\eqn\Tinvariance $$
For algebras of type $A$ or $E$, the second term on the right hand side is
always an integer, or can be chosen integer:
if $N$ is odd, $r$ is defined modulo $N$ and hence can always be chosen even
(provided one makes the same choice also in the third term), whereas if
$N$ is even by inspection one sees that $m$ must be even as well in order
for $m\rho$ to be an element of the root lattice, and hence
$mr/2 \in \Zbf$. Then the only threat to
$T$-invariance is the last term, $({r\over2} ml-1)ml NQ(a)Q(a)$. This is
an integer for any $a$ if and only if $({r\over2} ml -1)ml = 0 \mod N$.

Now we will determine the solutions to the three conditions (a), (b) and
(c) formulated above. Any solution to these conditions will be a positive
modular invariant of automorphism type, that can be obtained both
from Galois symmetry as well as from simple currents.

Consider first $E_6$. Condition (b) is trivial, so that $m$ has to satisfy (a)
$m(k+12)+1 \neq 0 \mod 3$, \ie\ $km+1 \neq 0 \mod 3$,
and (c) $(km -1) m = 0 \mod 3 $. We
may assume that $m\neq 0$ to avoid the trivial Galois scaling. Then
both conditions are satisfied if and only if $km = 1 \mod 3$.  There is
always a solution for $m$, namely $m=k\mod3$, unless $k$ is a multiple of 3.

Next consider $E_7$. Now $m$ has to be even in order that $m\rho$ is a root,
and this only allows the trivial solution $m=0$.

For $A_n$ the problem is a bit more complicated. As $T$-invariance
must hold for any charge $Q(a)$ it is clearly  sufficient to consider
$Q(a)={1\over N}$. Several cases have to be distinguished. We start
with odd $N=n+1$. Then condition (b) is automatically satisfied. For even
level  $k=2j$ the other two conditions read
$$\eqalign{&({\rm a})\quad \GCD(2jm + 1, N) = 1 \,, \cr
           &({\rm c})\quad (j m + 1 )m = 0 \mod N  \,.} \eqn\Conditions $$
The solution of the second equation depends crucially on the common
factors of $j$ and $N$. It is easy to see that if $j$ and $N$ have a common
factor $p$, then $m$ is divisible by $p$ as many times as $N$. In particular,
if $N=p^\ell$ and $j$ contains a factor $p$, then the only solution is the
trivial one. To remove common factors, write $j=j'q_a$, $m=m'q_b$ and
$N=N'q_b$, where $q_a$ is the greatest common divisor of $j$ and $N$,
and $q_b$ consists of all the prime factors of $q_a$ to the power with which
they appear in $N$. Now the second equation becomes
$$  (j' q_a m' q_b + 1 ) m' = 0 \mod N'   \ . \eqn\SecondEq $$
Now we know that $N'$ has no factors in common with $j'$, $q_a$ or $q_b$,
and hence we can find a $m'$ for which the first factor
vanishes $\mod N'$. This solution $m'$ is non-trivial provided $N' \neq 1$; if
$N'=1$ the solution is $m'=1$ (or 0), \ie\ $m=0 \mod N$.

The solution $m'$ has no factors in common with $N'$. Hence
we may write $2jm+1 = jm+(jm+1)=jm \mod N' = j' q_a m' q_b \mod N'$, so
that we see that $2jm+1$ and $N'$ have no common factors.
Furthermore $2jm+1$ and
$q_b$ have no common factors, since $m$ has a factor $q_b$.
Hence $2jm+1$ has no common factors with $N=N'q_b$, and therefore the
first equation is also satisfied.

In addition to the solution described here, \SecondEq\ may have additional
solutions with $m'$ and $N'$ having a common factor. It is again easy to see
that if $m'$ contains any such prime factor, it must contain it
with the same power
with which it occurs in $N'$. Let us denote the total common factor as
$p_b$, which is in general a product of several prime factors.
Then the second equation reads
$$ (j'q_a m'' p_b q_b + 1 ) m'' =  0 \mod N'' \ , \eqn\FinalAnswer $$
where $m'=m'' p_b$ and $N'=N'' p_b$.
We now look for solutions where $m''$ and $N''$ have no further common
factors. Such a solution does indeed exist, since the coefficient of $m''$ has
no factors in common with $N''$.  To show that the first condition is
also satisfied one proceeds exactly as in the foregoing paragraph.

When $N$ is odd and $k$ is also odd, we choose the even monodromy
parameter $r=k+N$, and define $j={k+N\over 2}$. The rest of the discussion
is then exactly as before.

If $N$ is even condition (b) implies that $m$ must be even as well, and
condition (c) becomes
$(km/2+1)m=0 \mod N$, or, writing $m=2t$, $N=2p$, $(kt+1)t=0 \mod p$.
Condition (a) reads $\GCD(km+1,N)=1$, which is equivalent to
$\GCD(2kt+1,p)=1$.  Now we have succeeded in bringing the conditions
in exactly the same form as \Conditions, and we can read
off the solutions almost directly. The only slight difference is that above
$N$ was odd, whereas here $p$ can be odd or even. However, the value of
$N$ did not play any r\^ole anywhere in the discussion {\it following}
\Conditions\ (it {\it was} used to derive \Conditions, though), and
hence everything does indeed go through.

If the algebra is $D_n$, $n$ odd, then we have to distinguish two cases. If
$n=1 \mod 4$, then condition (b) is trivially satisfied, and condition (a)
reads
$$           ({\rm a})\quad \GCD(m(k+2n-2)+1,4) = 1 \ , $$
from which we conclude that $mk$
(and hence $mr=3mk$) must be even, so that just as for $A_n$
and $E_n$ the second term on the right hand side of \Tinvariance\ plays
no r\^ole. Condition (c) thus reduces to
$$          ({\rm c})\quad  - ({r\over2}(mn)^2-mn) = 0 \mod 4\ ,     $$
with $k$ satisfying $3nk=r \mod 8$, or what is the same, $nk=3r \mod 8$.
To substitute this we multiply
the first argument of (a) with $n$, which does not affect this condition.
Afterwards we use that $n=1\mod4$, and then the conditions simplify to
 $$\eqalign{ &({\rm a})\quad  \GCD(3mr+1,4) = 1  \ ,\cr
             &({\rm c})\quad  - ({r\over2}\,m^2-m) = 0 \mod 4\ . \cr}$$
If $r$ is even, $r=2j$, condition (c) then
reduces to $jm^2-m=0\mod 4$. This clearly has a non-trivial solution if
$j$ is odd (then $m$ is odd), but only trivial solutions if $j$ is even.
If $r$ is odd the only solution to both equations is $m=2$.

If $n=3 \mod 4$ this argument goes through in much the same way,
but now solutions for odd $m$ are eliminated by condition (b).

\subsection{Automorphisms from fractional spin simple currents}\Mskip

Nearly all these results can be summarized as follows. Define $\tilde N = N$ if
$N$ is odd, $\tilde N = N/2$ if $N$ is even. Decompose $\tilde N$ into
prime factors, $\tilde N = p_1^{n_1} \ldots p_l^{n_l}$. Then the
set of solutions $m$ consists of all integers of the form $m=m''
{N\over \tilde N} p_1^{k_1} \ldots p_l^{k_l}$,
where $k_i = n_i$ if the monodromy parameter $r$ is divisible by $p_i$,
and $k_i = 0$ or $k_i = n_i$ otherwise.  The solutions
are thus labelled by all combinations of distinct prime factors of
$\tilde N$ that are not factors of $r$. The parameter $m''$ for each solution
in this set is the unique solution of the equation
$$ {1\over2} rlm''(p_1^{k_1}\ldots p_l^{k_l}) = 1 \mod  N''        \ , $$
where $N''={ \tilde N \over p_1^{k_1} \ldots p_l^{k_l} }$, and $r$ chosen
even if $N$ is odd.  These automorphism invariants have both a
Galois interpretation and a simple current interpretation:
they can be generated by the Galois
scaling $m(k+g)+1$ or alternatively by the fractional spin
simple current $J^{m}$.

These are precisely all the pure automorphisms generated by single
simple currents $K=J^m$ of fractional spin which have a ``square root", \ie\
for which there exists a simple current $K'$ such that $(K')^2=K$. Such a
square root exists always if $K$ has odd order, but if $K$ has even order it
must be an even power $m$ of the basic simple current $J$.  The
condition on the common factors of $r$ and $N$ has a simple
interpretation in terms of simple currents: If it is not
satisfied, then there are integral spin currents on the orbit of $J$. If one
constructs the simple current invariant associated with $J$
these currents extend the chiral algebra, so that one does
not get a pure automorphism invariant.

The condition that $K$ must have a square root is a familiar one:
in \ScYa\ the same condition appeared as a requirement that an invariant
can be obtained by a simple left-right symmetric orbifold-like construction
with ``twist operator" $L \bar L^c$.
If $K$ does not have a square root and $r$ is even, then there are additional
invariants,  which were described in \ScYa\ and derived in
\ScYb. Recently in \SchK\ it was observed that these invariants could
be described as orbifolds with discrete torsion. It is quite interesting
that precisely these discrete torsion invariants are missing from the
list of Galois invariants.

There is one exception, namely the automorphism invariants
of $D_{4l+1}$ at level $2j$, which are Galois invariants even though
they violate the foregoing empirical rule:
In this case $\tilde N = 2$, which is a factor of $r$.
Indeed, they are generated by the current $J_s$ (or $J_c$) which does
not have a square root.
Technically the reason for the existence of this extra solution is that this
is the only simply laced algebra with $\rho$ lying on the root lattice but
$N$ even.

\subsection{Automorphisms from integer spin simple currents}\Mskip

Finally, we have to return to the case $D_n$, $n$ even.
Since $M=2$ in this case, the only potentially
interesting solution is $m=1$. Hence $\Q$ is equivalent to $Q$ if and only
if $\rho$ is on the root lattice, which is true if and only if
$n = 0 \mod 4$. It is straightforward to derive the analogue of  \GaloisSimple:
$$ \sigma S_{ab} = S_{J_s^{2mQ_s(a)}J_c^{2mQ_c(a)}J_v^{(n-2)mQ_v(a)} a \,,\, b}
                            \ .  $$
(Since the three currents and charges are dependent this is a somewhat
redundant notation.)    The solution $m=1$ satisfies condition (a)
if and only if the level is even.  This
implies immediately that all three currents $J_s$, $J_v$ and $J_c$ have
integer spin, and we can write the transformation of $S$ in the following
symmetric way:
 $$ \sigma S_{ab} = S_{J_s^{2Q_s(a)}J_c^{2Q_c(a)}J_v^{2Q_v(a)} a \,,\, b} \ .$$
Since $Q_s+Q_c+Q_v=0\mod 1$ for any weight $a$, at least one of the charges,
say
$Q_v$, must vanish. Then $Q_s=Q_c \mod 1 $, and the field $a$ is transformed
to $J_s^{2Q_s(a)}J_c^{2Q_s(a)} a = J_v^{2Q_s(a)} a$. Since $J_v$ has
integral spin and $Q_v(a) = 0$, this field has the same conformal weight
as $a$, and hence $T$-invariance is respected. Due to the symmetry in
$s,c$ and $v$ the same is true for any other field as well. Thus we do
find an infinite series of modular invariant partition functions.  These are
automorphism invariants, again with both a Galois and a simple
current interpretation, although this time they are due to simple
currents of integer spin. Invariants of this type have been described
before in \Scha.

\subsection{Chiral Algebra Extensions}\Mskip

Now we will
examine what happens if we relax condition (b), \ie\ we will consider
the case that the replacement of $\Q$ by $Q$ leads to a different
answer. This obviously requires that $\rho$ is not on the
root lattice, and that the extra terms  \ExtraTerms\
are non-integral for some values of $Q$. The latter is true if $m$ is odd,
or if the algebra is $B_n, n$ odd, and $m=2$ ($\tilde m = 1$).
Now we can write (omitting for the moment the case $D_n, n$ even)
$$ \eqalign {\sigma S_{ab}
&= \eE^{-2\pi\ii l m N [Q(a) + {1\over2} ]
                              [Q(b)+ {1\over 2}] } S_{ab} \cr
&= \eE^{-\pi\ii l m N [Q(a)+{1\over2}]} S_{J^{-lmN[Q(a)+{1\over2}]}
   a\,,\,b} \cr} \eqn\GaloisExtension $$
instead of \GaloisCharge.
As before, a similar formula holds also for $B_n$, $n$ odd, with
$m$ replaced by $\tilde m = {\half m}$.

Since $mlNQ(a)$ is always integral and $N$ is even,
the exponential prefactor is in fact a sign, and the result may be written as
$$ \sigma S_{ab}= \eta(a)\, S_{J^{-lmN[Q(a)+{1\over2}]}a \,,\, b} \ .
\eqn\GalEx $$
Comparing this with \GaloisEps\ we find now that
$ S_{a,b}=\omega(a) S_{\tau a,b}\,, $
where $\omega$ is the product of the overall signs $\eta$ and
$\vareps$, and $\tau a=J^{-lmN[Q(a)+{1\over2}]}\dot{\sigma}a$.
Unitarity of $S$ now gives
$ \delta_{a,\tau a}=\sum_b S_{ \tau a,b} S_{ba}^*
 = \sum_b \omega(a) S_{ab} S_{ba}^* = \omega(a) ,$
which implies that $\omega=1$, \ie\ $\eta=\vareps$, and that $\tau$ is
the trivial map.

Also in this case the Galois transformation generates an automorphism
that lies within simple current orbits, and hence if it generates a
positive modular invariant, it must be a simple current invariant.
The identity is not
fixed in this case: it must thus be mapped to a simple current.
The candidate modular invariant has the form $P={\bf 1} + \eta(0) \Pi$,
where $\Pi$ is the matrix representing the transformation \GalEx.

Galois automorphisms of this type always have orbits with positive
and negative signs. A positive invariant can only be obtained if
the negative sign orbits are in fact fixed points of the Galois automorphism
(these should not be confused with fixed points of the simple current!).
One sees immediately from \GaloisExtension\ that the sign $\eta(a)$ is
opposite for fields of charge $Q(a)=0$ and $Q(a)={1\over N}$. Since the
former includes the identity we fix that sign to be positive. Hence the
orbits of charge ${1\over N}$ must be fixed points. This leads to the
condition
$$ -lmN[{1\over N} + {1\over 2}]  = 0 \mod N    \  , $$
or, writing $N=2N'$, $lm(N'+1)=0 \mod 2N'$. From this we conclude
that $N'$ must be odd and $lm$ must be a multiple of $N'=N/2$.

We are now in the familiar situation of an extension by a simple
current of order 2, and clearly $T$-invariance will then require this
current to have integral spin. The solutions can now easily be listed:
$$ \eqalign{&A_{4l+1}, \hbox{~level~} 4j \ \ \ \ \ (l, j\in \Zbf ) \,, \cr
            &B_{2l+1}, \hbox{~level~} 2j \ \ \ \ \ (l, j \in \Zbf )  \,,\cr
            &E_{7},\ \ \ \hbox{~~level~} 4j  \ \ \ \ \ (j\in \Zbf )  \,.\cr}$$

Now consider $D_{n}$ for even $n$. Then $\rho$ is not an element of the root
lattice, but a
vector weight if $n=2\mod 4$. Hence $Q_s(\rho)=Q_c(\rho)={1\over2}$ and
$Q_v(\rho) =0$.  The transformation of $S$ is now
$$ \sigma S_{ab}=\eE^{2\pi \ii [Q_s(a)+Q_c(a)] }S_{J_s^{2[Q_s(a)+{1\over2}]}
J_c^{2[Q_c(a)+{1\over2}]}a \,,\, b}\ , $$
where we set $m=1$, the only acceptable value. It is not hard to see
that the resulting $S$-invariant cannot be a positive one, since there do
exist wrong-sign Galois orbits that are not fixed points.

There are several simple current extensions that cannot be
obtained from Galois symmetry, at least not in the way described here.
Since we considered here only a single Galois scaling, only
Galois automorphisms of order 2 can give us a positive modular
invariant \FGSS\ (this is also true for the automorphism
invariants discussed earlier in this chapter, as one may
verify explicitly). Hence there is {\it a priori} no chance to obtain
extensions by more than one simple current.
However, some simple currents of order 2 are missing as well, namely
those generated by the current $J^{2l}$  of $A_{4l-1}$, the current
$J$ of $B_l, l$ even, and the currents $J_v$ of $D_l$ and $J_s, J_c$ of
$D_{2l}$, with levels chosen so that these currents have integer spin.
Note that the existence of a modular invariant of order two implies
the existence of a ``Galois-like" automorphism. This may suggest the
existence of some generalization of Galois symmetry that would also
explain those invariants.

\chapter{New Infinite Series}

In this chapter we will describe several infinite
series of  exceptional invariants that we obtained from Galois symmetry.
They occur for algebras of type $B$ and $D$ at level 2 and certain values
of the rank. Let us start the discussion with type $B$, which is
slightly simpler.

The new invariants occur for the algebras
$B_7$, $B_{10}$, $B_{16}$, $B_{17}$, $B_{19}$, $B_{22}$ etc., always
at level 2. The pattern of the relevant ranks $n$ becomes clear when we
consider the number $2n+1$, corresponding to the identity
$B_n=\SO(2n+1)$; namely, $2n+1$ must have at least two distinct prime factors.
For example, for $\SO(15)$ at level 2 we find the following three
non-diagonal modular invariants:
          \def\X{{\cal X}}
$$ \P_1=|\X_0+\X_1|^2
    +2\,(\, |\X_4|^2 +|\X_5|^2 +|\X_6|^2 +|\X_7|^2 +|\X_8|^2 +|\X_9|^2
    +|\X_{10}|^2 \,) \ , \hbox{~~~~~~~~} $$
$$ \P_2=|\X_0|^2 +|\X_1|^2 +|\X_2|^2 +|\X_3|^2 +|\X_5|^2 +|\X_6|^2 +|\X_8|^2
    + (\X_4\X_9^c + \X_7\X_{10}^c + \hbox{\rm c.c.} ) \ , $$
$$ \P_3= |\X_0+\X_1|^2 + | \X_4 + \X_9|^2 + | \X_7 + \X_{10}|^2
    + 2\,(\, |\X_5|^2 +|\X_6|^2 +|\X_8|^2 \,) \ . \hbox{~~~~~~~~~~~~~~} $$
Here the labels $i=1,2\ldots 10$ of $\X_i$ denote the following
representations:
$$\eqalign{
0 : \ \ \ \ \  &(0,0,0,0,0,0,0) \qquad\quad \ \
6 : \ \ \ \ \ (0,0,0,0,1,0,0) \cr
1 : \ \ \ \ \  &(2,0,0,0,0,0,0) \qquad\quad \ \
7 : \ \ \ \ \ (0,0,0,1,0,0,0) \cr
2 : \ \ \ \ \  &(0,0,0,0,0,0,1) \qquad\quad \ \
8 : \ \ \ \ \ (0,0,1,0,0,0,0) \cr
3 : \ \ \ \ \  &(1,0,0,0,0,0,1)\qquad\quad \ \
9 : \ \ \ \ \ (0,1,0,0,0,0,0) \cr
4 : \ \ \ \ \  &(0,0,0,0,0,0,2) \qquad\quad \,
10 : \ \ \ \ \ (1,0,0,0,0,0,0) \cr
5 : \ \ \ \ \  &(0,0,0,0,0,1,0) \cr} $$
The first of these invariants is not new: it corresponds to the conformal
embedding $\SO(15) \subset \SU(15)$. The fields $i=4 \ldots 10$ are fixed
points, each of which is resolved into two distinct
complex conjugate fields in the extended algebra. In $\SU(15)$ the two
fields originating from the $\SO(15)$ field $i$ are the antisymmetric
tensor representations $[4+i]$ and $[11-i]$. The invariant $\P_1$ is in fact
an integer spin simple current invariant. The other two $B_7$ invariants are
manifestly not simple current invariants.

The second $B_7$ invariant is new, as far as we know, and  can be
explained in the following way. The algebra $A_{14}$ at level 1 has
three distinct automorphism invariants which are generated by the
simple currents $J$, $J^3$ and $J^5$. They read
$$ \sum_{i=0}^{14} \X_i \X_{-i} \,, \quad\qquad 
 \sum_{i=0}^{14} \X_i \X_{-11i}  \,, \quad\qquad 
 \sum_{i=0}^{14} \X_i \X_{-4i} \,,  $$
respectively, where the labels are defined modulo 15.
The first one is equal to the charge conjugation invariant, and the last one
is the ``product" of the first two. The existence of an $A_{14,1}$ automorphism
implies relations among the matrix elements of the modular matrix $S$ of that
algebra.
Owing to the existence of the conformal embedding $B_{7,2}\subset A_{14,1}$,
these matrix elements are related to those of $B_{7,2}$.
The precise relation is
$$\eqalign{&S_{00}[A_{14,1}] = 2 S_{00}[B_{7,2}] \ ,
 \hbox{~~~~~~~~~~~~~~~~~~~~~~~~~~~~~~~~~~~~} \cr
&S_{0,4+i}[A_{14,1}]=S_{ 0,11-i}[A_{14,1}]=S_{0,i}[B_{7,2}] \ ,
 \hbox{~~~~~~~~~~~~~} \cr
&S_{4+i,4+j}[A_{14,1}]=S_{11-i,11-j}[A_{14,1}]\cr
=\,&S^*_{4+i,11-j}[A_{14,1}]
   =S^*_{11-i,4+j}[A_{14,1}]=\half S_{ij}[B_{7,2}] + \ii \Sigma_{ij}\ . \cr}
  $$
Here $\Sigma$ denotes the fixed point resolution matrix. The first
automorphism, charge conjugation, just sends $i$ to $-i$ and hence acts
trivially on the $B_{7,2}$ fields. The other two $\SU(15)$ automorphisms
interchange the $B_{7,2}$ fields $(4,9)$
and $(7,10)$, leaving $5,6$ and $8$ fixed (in addition one gets relations
from the imaginary part on the matrix elements of $\Sigma$). This implies
relations  like $S_{0,4}=S_{0,9}$ and $S_{4,7}=S_{9,10}$ for the $B_{7,2}$
matrix elements.  All these relations hold also if the label 0 is replaced
by 1, but we do not get any relations for matrix elements involving the fields
that are projected out, \ie\ the fields $2$ and $3$.
In the general case, the absence of relations involving fields that get
projected out implies that the automorphisms of an algebra g do not
lead to automorphims for a conformal subalgebra ${\rm h}\subset{\rm g}$.
The present case is an exception, since all the fields
on which the automorphism acts (and in fact all the fields with labels
$4 , \ldots, 10$) are fixed points of the $B_{7,2}$ simple current that
extends the algebra. Then the matrix elements $S_{2,i}$ and $S_{3,i}$
vanish for $i=4,\ldots,10$ and we need no further relations among them.

This explains the presence of the
second invariant listed above. The third one is a linear
combination of the foregoing ones and the diagonal invariant:
$\P_3=\P_1+\P_2-{\bf1}$. This
is a remarkable invariant: it looks like a normal extension by a spin 1
current, but it does not follow from any conformal embedding. The only
conformal embedding of $B_7$ at level 2 is in $\SU(15)$, and the
corresponding invariant is $\P_1$, not $\P_3$. This implies in particular that
there cannot exist any conformal field theory corresponding to the
modular invariant $\P_3$! In fact, it is not even possible to write down
a fusion algebra for this invariant, because there does not exist
a fixed point resolution matrix. In \ScYe\ another example of this kind
was described, although that theory was unphysical for a somewhat
different reason.

The existence of $\P_3$ can also be seen as a consequence of
the closure of the set of   Galois automorphims. Each Galois modular
invariant, automorphism invariants as well as chiral algebra
extensions, originates from a Galois symmetry of $S$, which acts on the
fields as a permutation accompanied by sign flips.  For the ``chiral
extension" $\P_3$ this Galois automorphism is represented by
the matrix $\P_3-\one$. This set of Galois automorphisms will always close
as a group. Indeed, the automorphism
underlying  $\P_3$ is simply the product of that of $\P_1$ and $\P_2$.

By the same arguments there will be pure automorphism invariants
for $B_{n,2}$ whenever $2n+1$ contains at least two different prime factors.
The spin-1 extension always involves an identity block plus $n$ fixed points
that yield each two $\SU(2n+1)$ level 1 fields (this is true since all
non-trivial representations of $\SU(2n+1)$ are complex).
If there is only one prime factor the only automorphism is charge conjugation,
which acts trivially. When there are $K$ different prime factors there are
$2^K$ distinct pure Galois automorphisms for $\SU(2n+1)$ at level 1, including
the identity and the charge conjugation invariant. When ``projected down"
to $B_{n,2}$ these are related in pairs by charge conjugation, and
we expect therefore
$2^{K-1}$ distinct $B_{n,2}$ modular invariants of  automorphism type. In
addition there is of course the invariant corresponding to the
conformal embedding in $\SU(2n+1)$ itself.
In combination with the $2^{K-1}-1$ {\it non-trivial} automorphisms
this extension gives rise to as many other invariants that look like
conformal embeddings, but actually do not correspond to a consistent
conformal field theory.

How does this come out in terms of Galois symmetry?
First of all the spin-1 extension of the conformal embedding is in fact
a simple current extension, and we have seen in the previous chapter that
it follows from Galois symmetry only for $B_n$ with $n$ odd.  If $n$ is odd
the Galois periodicity is $4(2n+1)$ for $B_{n,2}$ and $2(n+1)(2n+1)$ for
$A_{2n,1}$. Hence the cyclotomic field of the former is contained in that
of the latter, so that all Galois transformations of $A_{2n,1}$ have a
well-defined action on the modular matrix $S$ of $B_{n,2}$. In this case
we may thus
expect $2^K$ distinct Galois modular invariants, including the identity and
the unphysical invariants described above.
If $n$ is even the Galois periodicities are respectively
$2(2n+1)$ and $2(n+1)(2n+1)$, so that also in this case all Galois
transformations are well-defined on $B_n$. But due to the fact that
the simple current invariant is not a Galois invariant,
we get only half the number of invariants now, namely $2^{K-1}$.

For $n$ odd the  $\SU(2n+1)$ simple
current automorphisms are mapped to two $B_n$ modular invariants: one
physical automorphism and one chiral extension, which
(except for the one originating from the diagonal invariant, \ie\ the
conformal embedding invariant) is unphysical. For $n$ even each $\SU(2n+1)$
automorphism is mapped to just one $B_n$ invariant. The diagonal
invariant is mapped to the diagonal one of $B_n$, but it turns out that
the non-trivial automorphisms are mapped to either a pure automorphism
or an unphysical chiral extension, in such a way that the closure of
the set of Galois automorphisms is respected.

Now consider algebras of type $D$. Again the crucial ingredient is
the conformal embedding $\SO(2n)_2 \subset \SU(2n)_1$. In terms of
$D_n$ fields the $\SU(2n)$ characters are built as follows: The identity
character is the combination $\X_0 +\X_v$ and the antisymmetric tensor
$[n]$ has a character equal to $\X_s + \X_c$. All other $\SU(2n)$
representations are complex, and each pair of complex conjugate
representations arises from a resolved fixed point of the vector current of
$D_n$.  Even though $D_n$ has complex representations itself for $n$ odd, these
get projected out, and all the non-real contributions to
the $\SU(n)$ modular matrix $S$ arise from fixed point resolution.

The center of the $\SU(2n)$ \wzwt\ is $\Zbf_{2n}$, but the
`effective center' (in
the terminology of \BeBT) is $\Zbf_n$. This means that only the simple current
$J^{2}$ of the $\SU(2n)$ theory yields non-trivial modular invariants, and
that the order $2n$ current $J$ may be ignored. It is easy to see that the
field $[n]$ has zero charge with respect to $J^2$, so that it is mapped
onto itself by any automorphism generated by powers of $J^2$.
This implies that, just as before, all $\SU(2n)$ simple current
automorphisms act non-trivially only on resolved fixed points, and hence can be
`projected down' to $D_{n}$. If $n$ is prime, then
the only automorphism is equivalent to charge conjugation, and
hence it projects down to the trivial invariant. Hence just as before
we will get non-trivial $D_n$ automorphisms whenever $n$ contains at
least two distinct prime factors, where the prime is now allowed to be two. The
counting of invariants is the same as for $B_{(n-1)/2}$ above.
Again they come in pairs: an automorphism and an unphysical extension
by a spin-1 current.

All these invariants exist, but not all of them follow from Galois theory.
Just as for $B_n$, the automorphism invariants do, but the conformal
embedding invariant does not always follow. In fact, it never comes
out as a result of the scalings discussed in
the previous chapter. However, if $n=3 \mod 4$ the simple
current extension by the current $J_v$
is an exceptional Galois invariant only at level 2 (see the table in
the next chapter). In that case all the expected invariants
are Galois invariants. For all other values of $n$ only half of the
expected invariants are Galois invariants, and from each pair only
one member appears, either the automorphism or the unphysical extension.

There is still one interesting observation to be made here.
If there are just two distinct prime factors,
and $n=6 \mod 8$, then the extra invariant is an unphysical extension.
Remarkably, however, that extension is a simple current invariant. It
is equal to the extension by $J_v$, but it has additional terms
of the form $| \X_a + \X_b |^2$, where $a$ and $b$ are fields that
appear diagonally, as fixed points of order 2,
in the normal simple current invariant. The fields $a$ and $b$ are however
on the same orbit with respect to the current $J_s$, which makes this a simple
current invariant by definition. Nevertheless, it is not part of the
classification presented in \BeBT, because that
classification was obtained under a specific regularity condition
on the matrix $S$ that is not satisfied here (indeed, $D_{2n}$ at level 2
was explicitly mentioned as an exception in the appendix of \BeBT; the
reason for it being an exception is that all orbits except for the identity
field are fixed points of one or all currents). It also follows that this
simple current invariant cannot be obtained using orbifolds with discrete
torsion, unlike the simple current invariants within the classification \SchK.
Hence the fact that it is unphysical is not in contradiction with the
expectation that simple current invariants should normally be physical.

In the previous case the automorphism would be obtained by subtracting
the normal spin-1 extension, and adding the identity matrix. Clearly the
resulting automorphism is not really exceptional, but is simply
the automorphism generated by the spinor simple current $J_s$ (or $J_c$, which
at level 2 gives the same result). The same happens if the rank is $2 \mod 8$,
except that in that case the automorphism comes out directly as a Galois
invariant. It is listed in the table in the next section.
To get really new automorphisms that are not simple current invariants
for $n=2 \mod 8$ or $n=6\mod 8$ one has to consider cases where $n$
contains three or more distinct prime factors.
Finally, if the rank is divisible by 4 the spinor currents
have integer spin, and do not interfere with the exceptional automorphisms
discussed in this chapter.

\chapter{Pure Galois Invariants}

Here we list all the remaining Galois invariants of simple WZW models, \ie\
not including those described in the previous chapters.
All these invariants are positive and result directly from a single
Galois automorphism of order 2.
Although the full Galois commutant was investigated, in all but one
case there is only a single
non-trivial orbit contributing (in terms of the formula \modinv\ this means
that
$f_0$ is used to get $P_{00}=1$, and apart from $f_0$ only one
other coefficient $f_{\sigma}$ is non-zero.)
The exception is the $E_8$-type invariant of $A_1$ at level 28, which can
also be interpreted as a combined simple current/Galois invariant, and which
is therefore included in the table in the next chapter.
The results are listed in the following table. The notation is as
follows:
\pointbegin
CE: Conformal embedding.
\point
$S(J)$: Simple current invariant. The argument of $S$ is the simple current
responsible for the invariant.
\point
RLD: Rank-Level Dual. The $S$-matrices of $\SU(N)_k,\ \SO(N)_k$
and $C_{n,k}$ are related to those of respectively $\SU(k)_N,\ \SO(k)_N$ and
$C_{k,n}$ by rank-level duality.  One might expect that Galois
transformations of one matrix are mapped to similar transformations of the
other. The relation is not quite that straightforward however, and we will not
examine the details here. The results clearly respect this duality.
\point
EA: Exceptional Automorphism. These are modular invariants of
pure automorphism type that are not due to simple currents.
The only invariants of this type known so far were found
in  \Ver, and appear also in the table.
\point
HSE: Higher Spin Extension, an extension of the chiral algebra by currents
of spin larger than 1 that are not simple currents. Some of these invariants
can be predicted using rank-level duality; all other known ones are related
to meromorphic $c=24$ theories \SchM.

\Mskip\noindent
Note that there are some simple current
invariants in this list. This is not in conflict with the results of
chapter \ChapInf, as we did not claim that the list given there was complete.
The scales of the
Galois transformations for which these simple current invariants are obtained
are interesting. For $A_{4m-1}$ and $D_{4m+3}$ these scales are
equal respectively to $(2m+1)(k+g)-1$ and $3(k+g)-1$. If the contribution
$-1$ were replaced by $+1$,
they would be of the kind discussed in chapter \ChapInf. In fact we can
write these scales as $(-1)[(2m-1)(k+g)+1]\mod 4m(k+g)$ and  $(-1)[(k+g)+1]
\mod 4(k+g)$, respectively, which shows that these Galois automorphisms are
nothing but the product of a scaling of the type
discussed in chapter \ChapInf\ and  charge conjugation.
It can be checked that without the charge conjugation one does not get
a positive invariant: certain fields are transformed to their charge conjugate
with a sign flip. After multiplying with the charge conjugation
automorphism these fields become fixed points. The scale factor for $C_{4m}$,
$4m+3$, is of the form $(k+g)+1$, but for $C_n$ the arguments of
chapter \ChapInf\ break down right from the start, so that no conclusions can
be
drawn for this case. For the other simple current invariants the scale factor
does not have the right form, and hence the arguments of chapter
\ChapInf\ simply do not apply.

\vfill\endpage 

\begintable
Algebra | level | Galois scaling | Type | Interpretation \cr
$A_2$  | 5  |  19 | Extension |  CE $\subset A_5$ \nr
$A_{4m-1}$ | 2 |  $8m^2+8m+1$ | Extension | $S(J^{2m})$; RLD of $A_{1,4m}$ \nr
$A_4$ | 3 |  11 | Extension | CE $\subset A_9$ \nr
$A_9$ | 2 |  31 | Extension | RLD of $A_{1,10}$ \nr
$C_{4m}$ |  1 | $4m+3$ | Extension | $S(J)$; RLD of $C_{1,4m}=A_{1,4m}\;$\nr
$D_{8m+2}$ | 2 |   $8m+1$ | Automorphism | $S(J_s)$ \nr
$D_{4m+3}$ | 2 |   $24m+17$ | Extension | $S(J_v)$ \nr
$D_{7}$ | 3 |   49 | Extension | HSE; RLD of $\SO(3)_{14} = A_{1,28}\;$\nr
$G_{2}$ | 3 |  8  | Extension | CE $\subset E_6$ \nr
$G_{2}$ | 4 | 5 | Automorphism | EA \nr
$G_{2}$ | 4 |  11  | Extension  | CE $\subset D_7$ \nr
$F_{4}$ | 3 |  5   | Extension | CE $\subset D_{13}$ \nr
$F_{4}$ | 3 |  11  | Automorphism | EA \nr
$E_{6}$ | 4 |  7  | Extension | HSE \nr
$E_{7}$ | 3 |  13  | Extension | HSE \endtable
\vskip 1.truecm

\chapter{Combination of Galois and Simple Current Symmetries}

In chapter \ChapInf\ we have discussed a large set of invariants for which
the Galois and simple current methods overlap. If they do not overlap,
it may be fruitful to combine them. To do so we first have to understand
how the orbit structures of both symmetries are interfering with
each other.
This can be seen by computing $\sigma S_{Ja,b}$. On the one hand, this
is equal to
$$  \sigma S_{Ja,b} = \epss(Ja)\, S_{\sigmA J a,b} \ . $$
On the other hand, it is equal to
$$ \eqalign{ \sigma[ \eE^{2\pi \ii Q(b)}\, S_{ab}]
 &= \eE^{2\pi \ii l Q(b)}\epss(a)\, S_{\sigmA a,b}\cr
  &=\epss(a)\, S_{J^l \sigmA a,b} \ .  \cr} $$
Here $l$ is the power to which $\sigma$ raises the generator of the
cyclotomic field. In the first step we used that the simple current
phase factor is contained in the field $M$, which follows from
$\eE^{2\pi\ii Q(b)}=S_{Ja,b}/S_{ab}\in M$.
Using unitarity of $S$ we then find that
$$ \eqalign{  \epss(Ja) &= \epss(a)\,, \cr \sigmA J &= J^l \sigmA \ . \cr } $$
Here $J$ denotes the permutation of the fields that is generated by
the simple current $J$. Since $l$ is prime with respect to the order of
the cyclotomic field, it is -- at least in the case of WZW models --
also prime wih respect to the order $N$ of the simple current. If $N=2$
this means that $l$ must be odd so that $J^l=J$, and hence
we conclude that $\sigmA$ and $J$ commute. For all other values
of $N$ they do not commute unless $l=1 \mod N$,
but at least it is true that $\sigmA$ maps
simple current orbits to simple current orbits, and furthermore it
respects the orbit length.

If $N=2$ the simple currents yield the  relation
$$S_{Ja,Jb} = \eE^{2\pi \ii( Q(a) + Q(b) + {r \over 2}) } S_{ab} \  $$
among matrix elements of $S$,
where $r$ is the monodromy parameter. If $r$ is even (which is
the case for simple currents of integer or half-integer spin) this relation
takes the form
$$ S_{ab} = \vareps(a) \vareps(b) S_{Ja,Jb} \ , $$
since the phase factors are in fact signs. This is precisely the form of
a Galois symmetry, as expressed in \GaloisSymmetry.
We can represent this symmetry in matrix notation as
$$ \Pi_J S \Pi_J = S \ , $$
where $\Pi_J=(\Pi_J)^{-1}$ is an orthogonal matrix that commutes with the
analogous matrices representing the Galois group.
Hence we can extend the Galois group by this transformation as explained
in chapter \ChapGal. Furthermore
if $r = 2\mod4$ the simple current invariant produced
by $J$ is a fusion rule automorphism that can also be used to extend
the Galois group.

We have not examined these extended Galois-like symmetries
systematically, but we will illustrate that new invariants can be found
by giving one example. Consider $A_1$ at level 10. One of the Galois
invariants (invariant under $S$ as well as $T$) is
$$ \P_1 = | \X_0 + \X_6 |^2 + | \X_4 + \X_{10} |^2
            + |\X_1 - \X_9 |^2  + 2 | \X_3 |^2 + 2 |\X_7|^2  \ ,  $$
where the indices are the highest weights (in the Dynkin basis).
The only problem with this invariant is that it is not positive. However, at
level 10 we also have the $D$-type invariant
$$ \P_2 = | \X_0 |^2 + (\X_1 \X_9^* + \X_3 \X_7^* + \hbox{\rm c.c.})
    + |\X_2|^2 + |\X_4|^2 + |\X_5|^2 + |\X_6|^2 + |\X_8|^2 + |\X_{10}|^2 \ , $$
which is a simple current automorphism.
If we now take the linear combination $$\P_1+\P_2-\one\,,$$ we get a positive
modular invariant which is in fact the well-known $E_6$-type invariant.

There is a second way of combining simple currents and Galois symmetries.
One can extend the chiral algebra of the
WZW model by integer spin simple currents. This projects out some of
the fields, so that the negative sign Galois orbits of some Galois
invariants are removed. It is essential that the Galois automorphisms
respect the simple current orbits, and that the
matrix elements of $S$ are constant on these orbits
for the fields that are not projected out.
The simple current extension has its own $S$-matrix which can be
derived partly from that of the original theory. If $N$ is prime this
matrix has the form \ScYe\ScYg
$$ \tilde S_{a_i,b_j} = {N_a N_b \over N} S_{ab}E_{ij} + \Sigma_{ab} F_{ij}
\ ,\eqn\FixedPointS $$
where $\tilde S$ is the new modular matrix and $S$ the original one, $N_a$ is
the orbit length of the field $a$ (it is a divisor of the simple current
order $N$, and hence either 1 or $N$),
and $i$ labels the resolved fixed points for those orbits
with $N_a < N$ (\ie\ $i=1,\ldots,{N \over N_a})$, and analogously for
$b$ and $j$. The matrix $E_{ij}$ is equal to 1 independent of $i$ and $j$,
and $F_{ij}=\delta_{ij} - {1\over N} E_{ij}$. Finally, the matrix
$\Sigma_{ab}$ is non-vanishing only for fixed point fields and cannot
be expressed in terms of $S$, or at least not in any known way, but it
is subject to severe constraints from the requirement of modular invariance.

All general considerations regarding Galois transformations can be
applied directly to this new $S$-matrix. Clearly the matrix elements
$S_{ab}$ which correspond to the primary
fields of the original theory that are not projected out
belong to a number field $M'$ which is contained in the number field $M$
of the original theory. While $E_{ij}$, $F_{ij}$ and $N_aN_b/N$ are all
rational and hence transform trivially under $\Gal(M'/\rationals)$, the
presence of the
matrix $\Sigma_{ab}$ in \FixedPointS\ may require this number field to be
extended to a field $\tilde M'\supset M'$ (a simple example is provided by
the $A_{1,4}$ \wzwt, which has a real matrix $S$, whereas the $S$-matrix of the
extended algebra $A_{2,1}$ is complex). Now because of the projections
$\tilde M'$ does not necessarily contain the original number field $M$;
however, at the possible price of redundancies we can consider
an even larger number field $\tilde M$ that contains
both $\tilde M'$ and $M$. When working with $\tilde M$, we do
not loose any of the Galois transformations that act non-trivially on
the surviving matrix elements $S_{ab}$.
Note that any element of $\Gal(\tilde M/M)$ acts trivially on $S_{ab}$
and hence induces a permutation which leaves non-fixed points invariant
and acts completely within the set of
primary fields into which a fixed point gets resolved.
Further, for any element of $\Gal(\tilde M/\rationals)$
the associated permutation must act on the labels $a,\,b$ in the same way
in both terms on the right hand side of \FixedPointS.
In particular, for any matrix element involving only non-fixed points
the action of a Galois transformation on $S$ already determines its action on
$\tilde S$, since the two matrix elements are equal up to a rational factor.
The same is true for all
matrix elements between fixed points and full orbits, since in that
case $\Sigma$ is absent, too. This is often already enough information to
determine the Galois orbits of the extended theory completely. The
transformations of the fixed point - fixed point elements of $\tilde S$ are
more
subtle, and in principle would require knowledge of the matrix $\Sigma$.
However, as already pointed out
any element of the Galois group must act on $\Sigma$ exactly as it does on
$S$. Although this still leaves undetermined
the action within the set of primary fields into which the relevant fixed point
is resolved, this limited information nevertheless can provide
useful additional information on the matrix $\Sigma$, whose determination
in general is a problem that is far from being solved.

Fortunately, as long as we are only interested in modular invariants of the
original theory, we may in fact ignore fixed point resolution completely.
By definition that issue is determined solely by $S$
(and $T$), and the precise form of $\Sigma$ should not matter.

We have performed a computer search for invariants of the type described
above, and obtained the following list.

\vfill\endpage

\vskip 1.truecm
\begintable
Algebra | level | Galois scaling | Simple current | Type | Interpretation \cr
$A_1$  | 10  | 7   | ~~~~$J$ $^{(\dagger)}$ | Extension |  CE $\subset B_2$ \nr
$A_1$  | 28  | 11  | $J$ | Extension |  CE $\subset G_2$ \nr
$A_2$ | 9 |  17 |  $J$ |Extension | CE $\subset E_6$ \nr
$A_2$ | 21 | ~~35$\,\times\,$53 $^{(*)}$ | $J$ |Extension| CE $\subset E_7$ \nr
$A_3$ | 8 |  7 | $J$  | Extension | CE $\subset D_{10}$ \nr
$A_7$ | 4 |  7 | $J^2$ | Extension | HSE; RLD of $A_{3,8}$ \nr
$A_7$ |  $4m+2$ |  $4m+11$   | $J^4$ | Aut $\times$ Ext | $S(J^2)$   \nr
$A_{27}$ |  $2$ |  71  | $J^{14}$ | Extension | HSE; RLD of $A_{1,28}$   \nr
$C_3$ | 4 |   7 | $J$ | Extension | CE $\subset B_{10}$ \nr
$C_4$ | 3 |   7 | $J$ | Extension | RLD of $C_{3,4}$ \nr
$D_{4m+2}$ | $4l$ |  $8m+4l+3$  | $J_s$ | Extension | $S(J_v)\times S(J_s)$ \nr
$D_{4}$ | 6 |  5  | $J_s,J_v$ | Extension | CE $\subset D_{14}$ \endtable
\nobreak\vskip .5 truecm\noindent
$(\dagger)$ This is a simple current of half-integer spin;
see the main text for details.\hfill\break
$(*)$ Invariant originating from a non-cyclic subgroup $\Zbf_2 \times \Zbf_2$
of the \gagr. \hfill\break\vskip 1.truecm
Note that this list contains a few infinite series of simple current
invariants.
Since they were inferred from a finite computer scan, the statement
that the series continues is a conjecture.
Presumably these series can also be derived by arguments similar to those
in chapter \ChapInf, but we have not pursued this.

We have in principle
just looked for invariants originating from single orbits, but there is one
exception, namely the modular invariant of $A_2$ at level 21. This invariant is
obtained as a sum over a $\Zbf_2 \times \Zbf_2$ subgroup of the \gagr\ that is
generated by the two scalings indicated in the table. Separately each of these
scalings yields an $S,T$ invariant with a few minus signs.

\chapter{Conclusions}

To conclude, let us make a rough comparison between the
various methods for constructing modular invariants that were mentioned
in the introduction. We will compare them on the basis of the following
aspects.

\item{\bullet}{\bf Generality\hfill\break}
A common property of simple currents and
Galois symmetry is that neither is {\it a priori} restricted to WZW models,
unlike all other methods. (In practice this is less important than it may seem,
since essentially all RCFT's we know are WZW models or
WZW-related coset theories.)

\item{\bullet}{\bf Positivity\hfill\break}
Most methods do not directly imply the existence of positive modular
invariants, but rather they yield generating elements
of the commutant of $S$ and $T$ that have to be linearly combined to
get a positive invariant; the exceptions are simple currents, conformal
embeddings and rank-level duality.

\item{\bullet}{\bf Existence of a CFT\hfill\break}
It should be emphasized that a positive modular invariant partition function
is only a necessary condition for a consistent conformal field theory. Most
methods do not guarantee that a conformal field theory exists. Exceptions
are conformal embeddings (the new CFT is itself a WZW model) and
probably simple current invariants, since the construction of the new theory
can be rephrased in orbifold language.  Clearly any construction that may
yield negative invariants cannot guarantee existence of the theory, and
this includes Galois invariants. Indeed, we found examples of positive
Galois modular invariants that cannot correspond to any sensible CFT.

\item{\bullet}{\bf Explicit construction\hfill\break}
Simple current invariants can be constructed easily and straightforwardly.
On the other
hand, the explicit construction of an invariant corresponding to a conformal
embedding is usually extremely tedious. Indeed, many of these invariants
are not known explicitly. The other methods fall somewhere between these
two extremes. The explicit construction of a Galois invariant is
straightforward
but requires long excursions through the Weyl group, as explained
in the appendix.

\item{\bullet}{\bf Classification \hfill\break}
All simple current invariants have been classified in \BeBe, \BeBT\ and \SchK,
under a mild regularity assumption for $S$, which, as we have
seen in chapter \ChapNew, is not always satisfied. The simple
currents of WZW models were classified in \Fuch.
All conformal embeddings have been classified in \multref\CoSua{\CoSub}.
All cases of rank-level duality are presumably known, but all other methods
mentioned in the introduction have only been applied to
a limited number of cases, without claims of completeness. Our results on
Galois invariants are based partly on computer searches (inevitably
restricted to low levels) and partly on rigorous derivations
(chapter \ChapInf). For the pure Galois invariants we expect our results to be
complete, but we have no proof.

To summarize,
we find that the Galois construction does not yield all solutions, but
also that it is not contained in any of the previously known methods. It
generates invariants of all known types. Most of the partition functions we
found were already known in the literature, but we did find several
new infinite series of pure automorphism invariants not due to simple currents.

In the course of this investigation we realized that the restriction
that the scaling be prime with respect to $M(k+g)$ can in fact
be dropped, at least for WZW models. This yields even more relations among
elements of $S$, which take the form of sum rules, and hence even more
information about modular invariants. These
transformations, which we call `Quasi-Galois' symmetries,
will be discussed in a forthcoming paper.

\bigskip
\appendix{}  

Here we describe in detail how Galois scalings are implemented
when the \cft\ in question is a \wzwt\ based on an untwisted affine Lie
algebra g at integral level $k$. Then the \gagr\
is a subgroup of $\zett_{M(k+g)}^*$, where $g$ is the dual
Coxeter number of the horizontal subalgebra $\bar{\rm g}$ of $g$
(\ie\ the subalgebra generated by the zero modes of g) and $M$
is the denominator of the metric on the weight space of $\bar{\rm g}$.

We label the primary fields by the shifted highest weight
$a$ with respect to the horizontal subalgebra $\bar{\rm g}$, which differs
from the ordinary highest weight by addition of
the Weyl vector $\rho$ of $\bar{\rm g}$. Thus $a$ is
an integrable highest weight of $g$ at level $k+g$, \ie\ the components $a^i$
of $a$ in the Dynkin basis satisfy
  $$  a^i\in\zett_{\geq0} \quad {\rm for\ }i=0,1,...\,,{\rm rank}(\bar{\rm g})
  \,, $$
where $a^0\equiv k+\DC-\sum_{i=1}^{{\rm rank}(\bar{\rm g})}\theta_i a^i$
with $\theta_i$ the dual Coxeter labels of g.
However, because of the shift not all such integrable weights belong to
primary fields, but only the strictly dominant integral weights, \ie\
the primary fields of the WZW theory correspond precisely to those weights $a$
which obey
  $$  a^i\in\zett_{>0} \quad {\rm for\ }i=0,1,...\,,{\rm rank}(\bar{\rm g})
  \,. \eqn\ihw$$

A Galois transformation
labelled by $\ell\in\zett_{M(k+g)}^*$ acts as the permutation \CoGa
  $$  \sigmAl(a)=\hat w(\ell a) \,. \eqn\what  $$
If we label the fields by the weights $a-\rho$ which are at level $k$,
this is rewritten as in \wzw.
That it is the shifted weight $a$ rather than $a-\rho$ that
is scaled is immediately clear from the formula \moms\ for
the modular matrix $S$.
In fact, it is possible to derive the formula \GaloisSymmetry\ directly
by scaling the row and column labels of $S$ by $\ell$ and $\ell^{-1}$,
respectively,
using \what. Galois symmetry is thus not required to derive this formula,
nor is it required to show that \modinv\ commutes with $S$.
Galois symmetry has however a general validity and is not restricted
to WZW models.

Substituting \what\  into the formula for WZW conformal weights one
easily obtains a condition for $T$-invariance, namely
$(\ell^2-1) = 0$ mod $2M(k+g)$ (or mod $M(k+g)$ if all integers
$M\,a\cdot a $ are even).
 Since $\ell$ has an inverse mod $M(k+g)$, it follows that
$\ell=\ell^{-1}$ mod $M(k+g)$, i.e.\ the order of the transformation must be
2, as is also true \FGSS\ for arbitrary \cfts.

Let us explain the prescription \what\ in more detail. First one
performs a dilatation of the shifted weight $a=(a^1,a^2,\ldots)$ by the
factor $\ell\in\zett_{M(k+g)}^*$.
Now the weight $\ell a$ does not necessarily satisfy \ihw, \ie\
does not necessarily correspond to a primary field.
If it does not, then the dilatation has to be supplemented by the horizontal
projection $\hat w\equiv \hat w_{(\ell;a)}$ of a suitable affine Weyl
transformation.  More precisely, to any arbitrary integral weight $b$
one can associate an affine Weyl transformation $\hat w$ such that either
$\hat w(b)$ satisfies \ihw, and in this case $\hat w$ is in fact
unique, or else such that $\hat w(b)$ obeys $(\hat w(b))^i=0$ for some
$i\in\{0,1,...\,,{\rm rank}(\bar{\rm g})\}$ (in the latter case $\hat w(b)$
lies on the boundary of the horizontal projection of the fundamental Weyl
chamber of g at level $k+g$). To construct the relevant
Weyl group element $\hat w$ for a given weight $b$ as a product of
fundamental Weyl reflections $w_{(l)}$ (i.e.\ reflections with respect to the
$l$th simple root of g), one may use the following algorithm.
Denote by $j_1\in\{0,1,...\,,{\rm rank}(\gb)\}$ the smallest integer such
that $b^{j_1}<0$, and consider instead of $b$ the Weyl-transformed weight
$\hat w_1(b)$ with $\hat w_1:=\hat w_{(j_1)}$; next
denote by $j_2$ the smallest integer such that $(\hat w_1(b))^{j_2}<0$,
and consider instead of $\hat w_1(b)$ the
weight $\hat w_2\hat w_1(b)$ with $\hat w_2:=\hat w_{(j_2)}$, and so on,
until one ends up with an weight
$\hat w_n\ldots\hat w_2\hat w_1(b)$ obeying \ihw, and then
$\hat w=\hat w_n\ldots\hat w_2\hat w_1$ is the unique Weyl group element
which does the job. (The presentation of an element $\hat w\in\hat W$
as a product of fundamental reflections is however not unique; the present
algorithm provides one specific presentation of this type, which is not
necessarily reduced in the sense that the number of fundamental reflections
is minimal.)

It is worth noting that
there is no guarantee that starting from an integral weight $b$ one gets
this way a weight satisfying \ihw, but in the case where $b$ is of the form
$b=\ell a$ with $a$ integrable and $\ell$ coprime with $r(k+g)$, the
algorithm does work. Here $r$ denotes the maximal absolute value of the
off-diagonal matrix elements of the Cartan matrix of $\gb$, i.e. $r=1$
if $\gb$ is simply laced, $r=2$ for the algebras of type $B$ and $C$ and
for $F_4$, and $r=3$ for $\gb=G_2$. (The property that
$\ell$ is coprime with $r(k+g)$ in particular holds whenever \what\ corresponds
to an element of the \gagr, and hence for Galois transformations the algorithm
works simultaneously for all primary fields of the theory.)
Namely, assume that for some choice of $a$ there is no choice of
$\hat w\in\hat W$ such that $\hat w(\ell a)$ obeys \ihw. This means that
any $\hat w(\ell a)$ lies on the boundary of some affine Weyl chamber, and
hence
the same is already true for the weight $\ell a$. Then there must exist some
non-trivial $\hat v\in\hat W$ which leaves $\ell a$ fixed, $\hat v(\ell a)=
\ell a$. Decomposing $\hat v$ into its finite Weyl group part $v\in W$ and its
translation part $(k+g)t$ (with $t$ an element of the coroot lattice of
$\bar{\rm g}$), this means that we have $\ell\,v(a)+(k+g)t=\ell a$, or in
other words,
  $$  \ell\,(a-v(a))=(k+g)\,t \,.  \eqn\ava $$
Now assume that $\ell$ is coprime with $r(k+g)$. This implies that there exists
integers $m,\,n$ such that $m\ell=nr(k+g)+1$. Multiplying \ava\ with $m$
then yields
  $$  a = v(a) + (k+g)[mt-nr\,(a-v(a))]  \,.  \eqn\avA $$
Since for any integral weight $a$ the weight $r(a-v(a))$ is an element of
the coroot lattice, the same is also true
for the expression in square brackets, and hence \avA\ states that the
weight $a$ stays fixed under some affine Weyl transformation. But $a$
satisfies \ihw, and hence the fact that $\hat W$ acts freely on such weights
implies that this Weyl transformation must be the identity. This
implies that $\hat v$ must be the identity as well. Thus for $\ell$
coprime with $r(k+g)$ the assumption that $\hat w(\ell a)$ is not integrable
leads to a contradiction.

In the general case where $b$ is not of the form $\ell a$ with $a$ subject
to \ihw\ and $\ell$ coprime with $r(k+g)$, the algorithm
described above still works unless at one of the intermediate steps one
of the Dynkin labels becomes zero, which means that the weight lies on the
boundary of the fundamental affine Weyl chamber. In the latter case any
Weyl image of this weight lies on the boundary of some affine Weyl chamber
as well, and hence we can never end up with a weight that satisfies \ihw, i.e.\
in the interior of the fundamental affine Weyl chamber.
It may also be remarked that one can speed up the algorithm considerably
using not the weight $b$ itself as a starting point, but rather
the weight $\tilde b=b+(k+g)t$ that is obtained from $b$ by such a Weyl
translation $(k+g)t$ for which the length of $\tilde b$ becomes minimal.

Finally, there is a
general formula for the sign $\epsilon_{\sigmal}^{}$, namely
  $$   \epsilon_{\sigmal}^{}(a)=\eta^{}_\ell\,{\rm sign} (w_{(\ell;a)}) \,,$$
i.e.\ the sign is just given by that
of the Weyl transformation $\hat w$, up to an overall sign $\eta_\ell$ that
only depends on $\sigmal$ \CoGa, but not on the individual highest weight $a$.
(Actually the cyclotomic field $\rationals(\zeta_{M(k+g)})$ whose \gagr\ is
$\zett_{M(k+g)}^*$ does not yet always contain the overall normalization
$\cal N$ that appears in the formula \moms\ for $S$, but rather sometimes
a slightly larger cyclotomic field must be used \CoGa.
However, the permutation $\dot\sigma$ of the primary fields that is induced
by a Galois scaling can already be read off the generalized quantum
dimensions, which do not depend on the normalization of $S$. The
correct Galois treatment of the normalization of $S$ just amounts to the
overall sign factor $\eta_\ell$, which is irrelevant for our purposes.)

\bigskip
\par \penalty-4000\vskip\chapterskip
   \spacecheck\referenceminspace \immediate\closeout\referencewrite
   \referenceopenfalse
   \line{\fourteenrm\hfil REFERENCES\hfil}\vskip\headskip
   \endlinechar=-1
   \input referenc.texauxil
   \endlinechar=13
   
\end